\pdfoutput=1

\documentclass[pre,twocolumn,showpacs,aps,floats,floatfix]{revtex4-1}

\usepackage{natbib}
\usepackage[pdftex]{graphicx}
\usepackage{amssymb,amsmath}
\usepackage{amsfonts}
\usepackage{mathrsfs}
\usepackage[english]{babel}
\usepackage{epsfig}
\usepackage{epstopdf}
\usepackage{babelbib}
\usepackage{hyperref}
\usepackage{color}
\usepackage{tcolorbox}
\usepackage{tabularx}
\usepackage{array}
\usepackage{colortbl}
\tcbuselibrary{skins}
\usepackage{comment}
\usepackage{array}
\usepackage{bm}
\usepackage{mathtools}
\usepackage[utf8]{inputenc}

\begin{document}

\title{Mechanical design of apertures and the infolding of pollen grain}

\author{An\v ze Bo\v zi\v c}
\affiliation{Department of Theoretical Physics, Jo\v zef Stefan Institute, Jamova  39, 1000 Ljubljana, Slovenia}
\author{Antonio \v Siber}
\affiliation{Institute of Physics, Bijeni\v{c}ka cesta 46, 10000 Zagreb, Croatia}
\email{antonio.siber@ifs.hr}

\date{\today}

\begin{abstract}
When pollen grains become exposed to the environment, they rapidly desiccate. To protect themselves until rehydration, the grains undergo characteristic infolding with the help of special structures in the grain wall---apertures---where the otherwise thick exine shell is absent or reduced in thickness. Recent theoretical studies have highlighted the importance of apertures for the elastic response and the folding of the grain. Experimental observations show that different pollen grains sharing the same number and type of apertures can nonetheless fold in quite diverse fashion. Using thin-shell theory of elasticity, we show how both the \emph{absolute} elastic properties of the pollen wall as well as the \emph{relative} elastic differences between the exine wall and the apertures play an important role in determining pollen folding upon desiccation. Focusing primarily on colpate pollen, we delineate the regions of pollen elastic parameters where desiccation leads to a regular, complete closing of all apertures and thus to an infolding which protects the grain against water loss. Phase diagrams of pollen folding pathways indicate that an increase in the number of apertures leads to a reduction of the region of elastic parameters where the apertures close in a regular fashion. The infolding also depends on the details of the aperture shape and size, and our study explains how the features of the mechanical design of apertures influence the pollen folding patterns. Understanding the mechanical principles behind pollen folding pathways should also prove useful for the design of the elastic response of artificial inhomogeneous shells.
\end{abstract}

\maketitle

\section*{Introduction}

Pollination is a crucial process in the life cycle of plants. For it to proceed, pollen grains must leave the anther, which exposes them to rapid desiccation as they cannot actively control their hydration status~\cite{Dumais2013,Hoekstra2002,Firon2012}. The grains thus require some sort of protective mechanism against desiccation in the period before they land on the stigma of a flower, where they germinate upon rehydration to complete the fertilization. The near-universal protective mechanism against desiccation in pollen during presentation and dispersal is {\em harmomegathy}: a characteristic infolding of the grain in response to a decreasing cellular volume upon dehydration~\cite{Wodehouse,Pollen,Payne1972,Franchi2011,Pacini2020}.

\begin{figure}[tb]
\centering
\includegraphics[width=0.9\columnwidth]{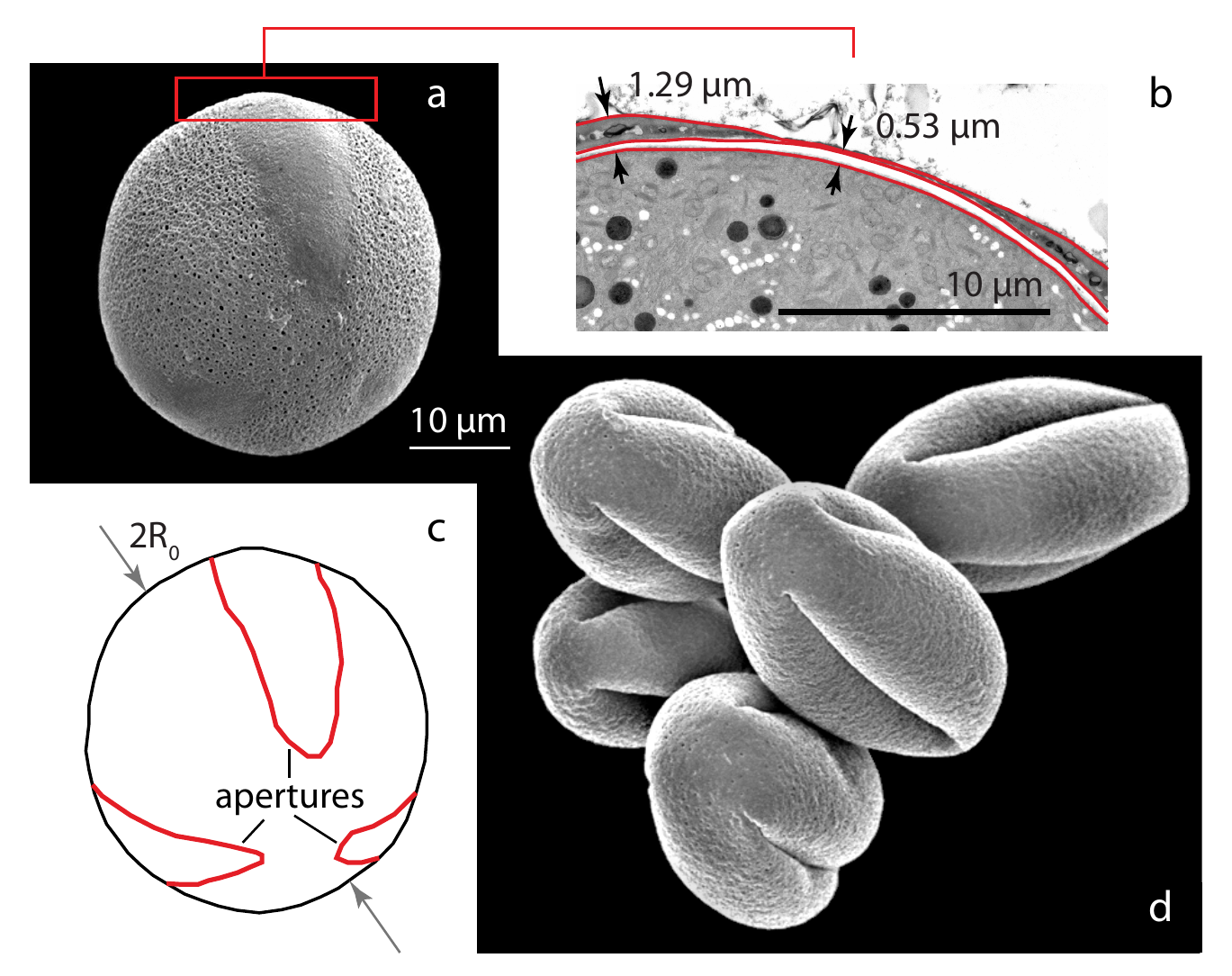}
\caption{Tricolpate pollen of {\em Betonica officinalis} (common hedgenettle) {\bf (a)}. Interapertural thickness of both intine and exine together {\bf (b)} is approximately $d\approx1.2$ $\mu$m, while the thickness of the intine in the apertural region (where the exine is absent) is approximately half of that value. The hydrated pollen shape in {\bf (a)} is spheroidal with a (mean) diameter of $2R_0 \approx 35$--$40$ $\mu$m {\bf (c)}. The pollen elongates upon desiccation when the apertures close and fold inwards {\bf (d)}. Pollen images by H.\ Halbritter and S.\ Ulrich~\cite{Betonica}, used with permission from the Society for the Promotion of Palynological Research in Austria.}
\label{fig:1}
\end{figure}

The shell of a pollen grain encapsulates the male plant genetic material it carries. It consists of two bio-polymeric layers: intine, a soft cellulosic interior layer, and exine, a hard exterior layer composed of sporopollenin and impermeable to water~\cite{Hoekstra2002,Franchi2011,Pollen}. The two layers are not homogeneous throughout the shell, and regions which significantly differ structurally and morphologically from the rest of the shell wall and where the exine layer is either absent or reduced in thickness are termed {\em apertures}~\cite{Franchi2011,Prieu2016,Pollen,Volkova2013,Pacini2020}. Not only do apertures play a role in the harmomegathic accommodation of grain volume changes, they also function as sites for water uptake and the initiation of pollen tubes~\cite{Pollen,Dajoz1991,Muller1979}. Despite these shared functions of apertures, their number, shape, and size vary greatly among and within pollen species~\cite{Pollen,PALDAT}. Most commonly, apertures occur in the form of a spherical lune (colpus or sulcus), a circular region (porus or ulcus), or a combination of the two (colporus). Monosulcate pollen with a single distal aperture is ancestral in angiosperms (flowering plants)~\cite{Walker1975,Furness2007,Furness2004}, and remains a distinctive feature of a large evolutionary line of monocots. In eudicots, a very large clade comprising approximately 75\% of the extant angiosperm species, pollen grains are most often characterized by {\em three} apertures~\cite{Furness2004}---see Fig.~\ref{fig:1}. In general, many aperture patterns can be observed in angiosperm pollen, and some of them may not easily fall in a predetermined category. Angiosperm pollen can also be inaperturate or omniaperturate, and evolutionary trends appear to favor an increasing number of apertures~\cite{Dajoz1991,Furness2004,Prieu2016}. Nonetheless, the vast majority of angiosperms produce pollen with one or three apertures~\cite{Albert2018,Nowicke1979,Khansari2012,Wang2018,PALDAT,Harley2004,Furness2007,Furness2004}.

During desiccation, apertures often fold inwards, and in many pollen species the edges of each aperture eventually touch each other and effectively close off, thus reducing the rate of water loss~\cite{Bolick1992,Matamoro2016,Katifori2010} [Fig.~\ref{fig:1}d]. The precise effects of the aperture number or shape on harmomegathic volume accommodation are, however, still not completely understood~\cite{Prieu2016,Pacini2020}. Empirical studies have shown that harmomegathy depends on the aperture pattern, but it also appears to be influenced by other characteristics such as the size or shape of the grain~\cite{Halbritter2004,Prieu2016}. Some flexibility of the exine wall is also required---since even pollen without apertures is able to accommodate volume changes to a degree~\cite{Muller1979}---and exine ornamentation and its elastic properties have been suggested to be involved in the process as well~\cite{Matamoro2016,Katifori2010}. Overall, the entire suite of structural adaptations of angiosperm pollen grain seems to be tailored to favor large-scale, mostly inextensional bending of its wall, allowing the apertures to fold inward and in this way reduce the rate of desiccation~\cite{Halbritter2004,Katifori2010,Couturier2013,Dumais2013}.

Depending on the species, pollen changes its shape during desiccation in surprisingly different ways, from a regular infolding to a seemingly random, irregular fashion---even if the type of exine ornamentation and the aperture condition appear to be almost identical~\cite{Hoekstra2002,Pollen,Halbritter2004}. Several theoretical studies have recently explored the role that apertures in pollen grains~\cite{Katifori2010,Couturier2013} or, more generally, local soft spots in elastic shells~\cite{Shim2012,Paulose2013,Lin2015} play in their folding pathways. Studies of instabilities and buckling in thin shells~\cite{Karman1939,Pogorelov,Landau,Siber2009,Vliegenthart2011} have shown that a perfectly homogeneous spherical shell exposed to uniform pressure or undergoing a change in volume will develop depressions in unpredictable positions due to the high symmetry of the problem~\cite{Vliegenthart2011,Goriely}. The buckling behavior can however be altered and guided by creating local weak spots in an otherwise uniform shell~\cite{Shim2012,Paulose2013,Lin2015}, and this elastic inhomogeneity of the shell can also be realized by varying its thickness~\cite{Vernizzi2011,Datta2012,Munglani2019}. Similar considerations can be applied to pollen grains~\cite{Bolick1981}, where the high sporopollenin content makes exine a very stiff material~\cite{Rowley2000,Liu2004,Qu2018} while the apertures can be seen as elastic soft spots in the pollen wall~\cite{Katifori2010}. Using a thin-shell elastic model of the pollen grain, Katifori et al.~\cite{Katifori2010} have shown that apertures contribute to harmomegathy by reducing the necessity of the pollen wall to significantly stretch and bend in order to accommodate volume changes, guiding the pollen to fold in a regular fashion by closing the apertures. While these models thus confirm that various morphological and structural adaptations enable adjustments of pollen grains to sudden volume changes by influencing their mechanical properties, not much is known about the precise nature of this relationship.

A detailed knowledge of the elastic properties of the grain and their spatial variation is required in order to formulate a model of pollen infolding: a soft spot in an overall soft shell will lead to a different folding pathway than the same soft spot in an otherwise hard shell. Yet the mechanical properties of pollen remain largely unexplored~\cite{Rowley2000,Liu2004,Qu2018}, and little is known about how these properties and their variation influence folding pathways of pollen grains. Our work addresses this important question by using an elastic model which is versatile enough to investigate different modes of pollen folding based on the estimated elastic properties of pollen walls. This allows us to explore phase diagrams of folding pathways with respect to the overall elastic properties of pollen walls and the degree of weakening provided by the apertures. In this way, we are able to distinguish between regions where the apertures successfully close and the (possibly irregular) pathways where they do not close properly. We also show how the aperture size, shape, and their number all play a role in determining the folding pathways of pollen grains.

\section*{Results}

\subsection*{Elasticity of pollen grains}

A significant amount of monadous pollen, shed as a single pollen grain, is approximately spherical in its hydrated form~\cite{PALDAT,Franchi2002} [Fig.~\ref{fig:1}]. For the purposes of the mechanical modelling of pollen grains, we effectively treat the entire interior of the grain as a liquid which acts only to constrain the total volume enclosed by the pollen wall. Such a model could also be formulated in terms of internal pressure imparted on the elastic spherical shell by the water inside it~\cite{Katifori2010}. Using a discrete elastic model (see Materials and Methods), we assign different microscopic elastic parameters (stretching parameter $\epsilon$ and bending parameter $\rho$) to the exine region of the pollen wall and to the apertures where the exine is either absent or thinned. Parameter $f<1$ is the ratio of the elastic parameters in the different regions and it sets the elastic difference between the pollen wall and the apertures---the smaller the value of $f$, the softer the apertural region compared to the exine wall. The overall elasticity of the pollen grain is on the other hand determined by the dimensionless F\"oppl-von K\'arm\'an (FvK) number $\tilde{\gamma}$, proportional both to the ratio of the stretching and bending parameters of the pollen wall and to the square of its radius $R_0^2$ (as described in Materials and Methods). It is important to stress that the FvK number $\tilde{\gamma}$ is a quantity which applies to the {\em entire} grain surface, both in the exine region and in the apertures, since the softness parameter $f$ cancels out in the ratio of the elastic parameters in the aperture regions. While $\tilde{\gamma}$ can be thus understood as a measure of the overall, global elasticity of the grain, $f$ indicates the extent of elastic inhomogeneity due to the presence of the apertures.

\subsection*{Phase diagrams of tricolpate pollen folding}

Desiccation and the corresponding volume change lead to infolding of pollen grains. We first study folding pathways of tricolpate pollen, both because triaperturate pollen is the most common in angiosperms and because colpi usually bring about a more regular aperture closing than some other aperture shapes, such as pores~\cite{Katifori2010,Payne1972}. We model the colpi as spherical lunes extending from pole to pole, whose size is given by their azimuthal opening angle $\phi_c$. In an $n$-colpate pollen, such apertures span a total angle of $\phi_{ap} = n \phi_c$ and represent $A_{ap}/A_0=n\phi_c/2\pi$ of the total grain area $A_0$. While it is difficult to precisely determine the total area that apertures cover in pollen grains, it can still be fairly reliably estimated from the microscopic images of pollen, where available~\cite{PALDAT,Pollination}. In tricolpate pollen, $A_{ap}/A_0$ typically ranges from $0.1$ to $0.4$, depending on the species~\cite{PALDAT}, and the value which we use as representative in this section is $A_{ap}/A_0=1/3$. In the next section, we address both the effect of varying the number and size of colpi on the folding pathways of pollen, and since colpi and circular pores can have different harmomegathic patterns~\cite{Payne1972,Katifori2010}, we also investigate the role of the aperture shape.

\begin{figure}[tb]
\centering
\includegraphics[width=\linewidth]{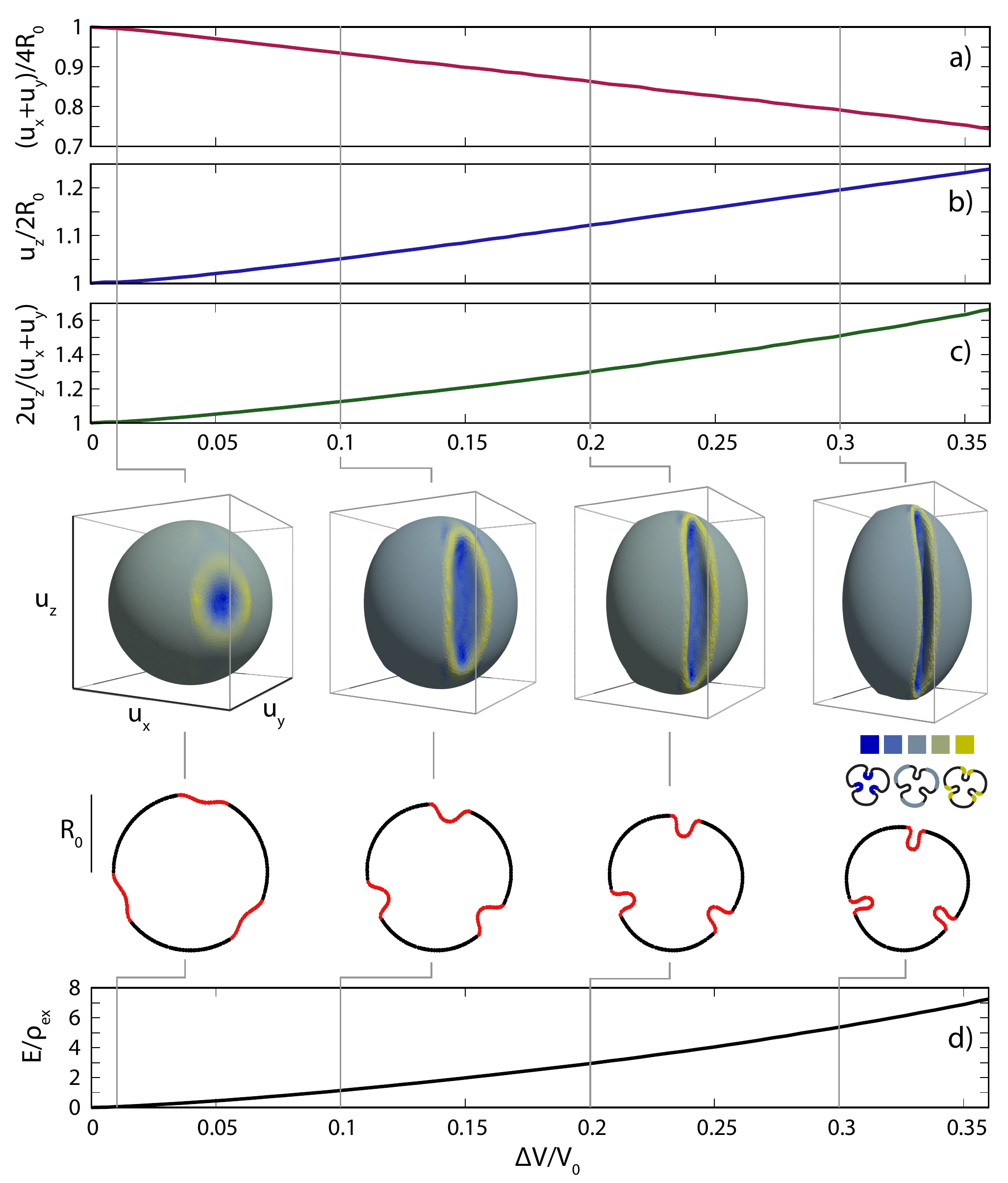}
\caption{Process of regular aperture closing during desiccation of tricolpate pollen with $\tilde{\gamma}=7000$, $f=0.01$, and $A_{ap}/A_0=1/3$. The plots show the change in different parameters upon a decrease in the relative volume $\Delta V/V_0$: the scaled average size of the bounding box in the equatorial plane, $(u_x+u_y)/4R_0$ {\bf (a)}; the scaled size of the bounding box along the $z$-axis, $u_z/2R_0$ {\bf (b)}; their ratio {\bf (c)}; and normalized elastic energy {\bf (d)}. Insets show 3D shapes in their bounding boxes (in perspective projection), and their equatorial cross-sections at different values of $\Delta V/V_0$. The cross-sections contain the points located in the ring around the equator of the mesh, with $z$-coordinates between $-0.04 R_0$ and $0.04 R_0$; the apertures are indicated in red. Colors in the 3D shapes indicate the mean surface curvature, the brightest yellow and the darkest blue in each individual shape corresponding to its largest and smallest curvature, respectively. In these shapes, the largest mean curvature (yellow) is positive and the smallest (blue) is negative.}
\label{fig:2}
\end{figure}

Figure~\ref{fig:2} shows the prototypical, regular infolding process of a tricolpate pollen grain with $\tilde{\gamma}=7000$ and $f=0.01$ as its volume decreases from the initial volume $V_0$ to about $V=0.65V_0$ (the volume difference is denoted by $\Delta V = V_0 - V$). The overall grain shape and its changes can be characterized by the dimensions of the minimal bounding box which contain it, $(u_x, u_y, u_z)$; for a fully hydrated spherical shape, $u_x=u_y=u_z=2 R_0$. Colpi are arranged equatorially, and the poles of the pollen grain, i.e., the points where the colpi meet, are positioned on the $z$-axis. As the infolding proceeds, the equatorial dimensions of the bounding box containing the pollen grain shrink (Fig.~\ref{fig:2}a) and the distance between the poles increases (Fig.~\ref{fig:2}b). The shape of the grain thus becomes prolate, which can also be quantified by the ratio of the sizes of the bounding box in the $z$-direction and in the equatorial plane (Fig.~\ref{fig:2}c). The regular infolding of the grain and its elongation along the $z$-axis require an increase in the elastic energy of the shape (Fig.~\ref{fig:2}d). The apertures close completely once a large enough reduction in volume is reached (in this case, $\Delta V/V_0\sim0.35$). Past this point, the edges of the apertures touch, the shell enters a self-collision regime and further reduction of the volume requires additional deformation of the exine. At the point of the aperture closing, the ratio of the pollen long and short axes exceeds $1.6$ in this case. A prolate shape is perhaps the most commonly observed feature in dry pollen grains---this is a direct consequence of a regular infolding in colpate grains, which can otherwise be perfectly spherical in their fully hydrated state, as is also the case in our calculations.

\begin{figure*}[t]
\centering
\includegraphics[width=1.5\columnwidth]{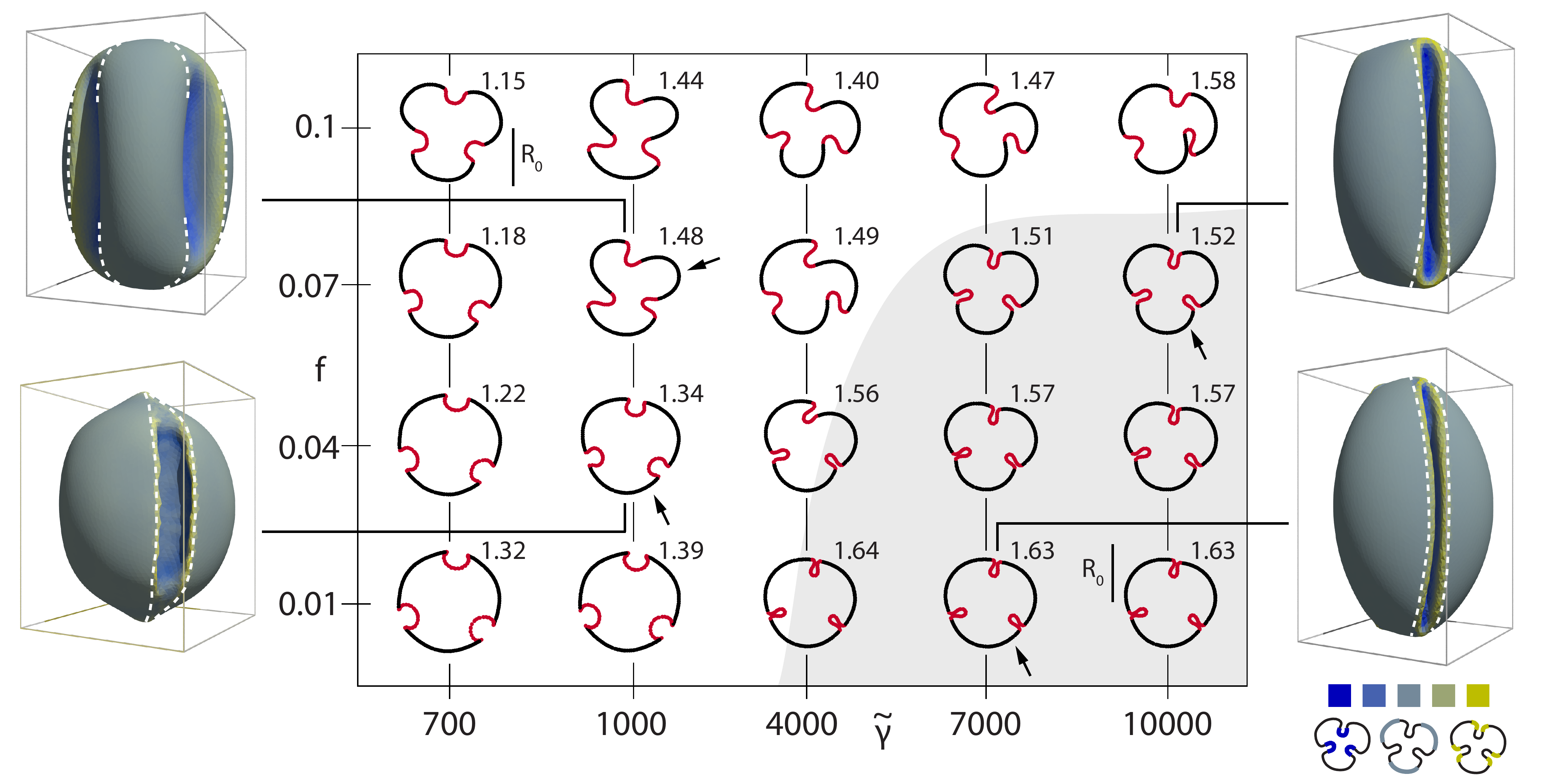}
\caption{Phase diagram of tricolpate pollen folding in the $(f,\tilde{\gamma})$ plane, showing equatorial cross-sections of pollen shapes. The relative change of volume upon desiccation is $\Delta V/V_0=0.35$ throughout, and the colpi (shown in red in the cross-sections) span one third of the total area of the pollen surface, $A_{ap}/A_0=1/3$. Numbers next to the pollen shapes denote the elongation of the bounding box of the shapes, $2u_z/(u_x+u_y)$, indicating the prolateness of the grain. Insets show 3D shapes of desiccated pollen in different parts of the phase diagram, where the arrows indicate the viewpoint with respect to their equatorial cross-section; the colors have the same meaning as in Fig.~\ref{fig:2} and the dashed white lines indicate the borders of the apertures. The shaded region of the phase diagram corresponds to the elastic parameters where the apertures close completely and nearly symmetrically, without asymmetric deformation of the exine.}
\label{fig:3}
\end{figure*}

While the apertures of a tricolpate pollen grain with $\tilde{\gamma}=7000$ and $f=0.01$ close completely once the volume of the grain is reduced by one third, such a reduction in the initial volume is in general not necessarily enough for the apertures to close. Whether this happens at a certain $A_{ap}/A_0$ depends both on the FvK number $\tilde{\gamma}$ and on the ratio of the elastic constants of the apertures and the exine $f$. Phase diagram of infolded shapes of tricolpate pollen in the $(f,\tilde{\gamma})$ plane, shown in Fig.~\ref{fig:3}, makes it clear that apertures close completely and in a regular fashion described in Fig.~\ref{fig:2} only in a restricted region of the phase diagram. Complete, regular or nearly regular closure of the three colpi requires a pollen grain with $\tilde{\gamma}\gtrsim3000$ and $f\lesssim0.07$. When the apertures are not soft enough compared to the exine (i.e., when $f$ is relatively large, $\sim0.1$), they do not function well as the regions which dominantly localize the infolding---the pollen grain folds in an asymmetric fashion, with only some of the three colpi closing completely. On the other hand, for sufficiently low values of the FvK number, $\tilde{\gamma}\lesssim1000$, the apertures infold symmetrically but do not close completely, even though they are soft compared to the exine.

\subsection*{Role of aperture shape, size, and number}

Inspection of folding pathways of tricolpate pollen points to the importance of the elastic properties of the entire pollen grain and the elastic inhomogeneities provided by the apertures---both of these determine whether or not the apertures close upon desiccation and the manner in which they close. However, at a fixed FvK number of the grain $\tilde{\gamma}$ and the softness of the apertures $f$, the pollen folding pathways are further determined by the shape, size, and number of the apertures.

In the results presented thus far, the colpi extended from pole to pole: this is an idealized case (see Fig.~\ref{fig:1}). Even a small reduction in the polar span of the apertures can change the infolded shape of the desiccated pollen, which typically becomes markedly more lobate, and further reduction of the polar span partially invalidates the function of the apertures, which do not close completely anymore. In addition, large exine regions flatten upon infolding. Eventually, the apertures completely fail and the grain buckles in an irregular fashion, as large intraapertural region become sunken so that the infolded shape resembles a cup. Such a shape is often found in dry inaperturate and porate pollen grains~\cite{PALDAT}. The progressive change of the aperture shape is addressed in SI Appendix, and these results are illustrated in SI Appendix, Fig.~S2. Variation of the area covered by the apertures, on the other hand, does not seem to influence much whether the apertures close completely or not. However, changing the aperture area (while retaining its shape) influences the amount of the volume reduction which can be achieved by the full closing of the apertures---the smaller the aperture area, the smaller the volume reduction upon closing (demonstrated in SI Appendix, Fig.~S3).

\begin{figure}[t]
\centering
\includegraphics[width=\columnwidth]{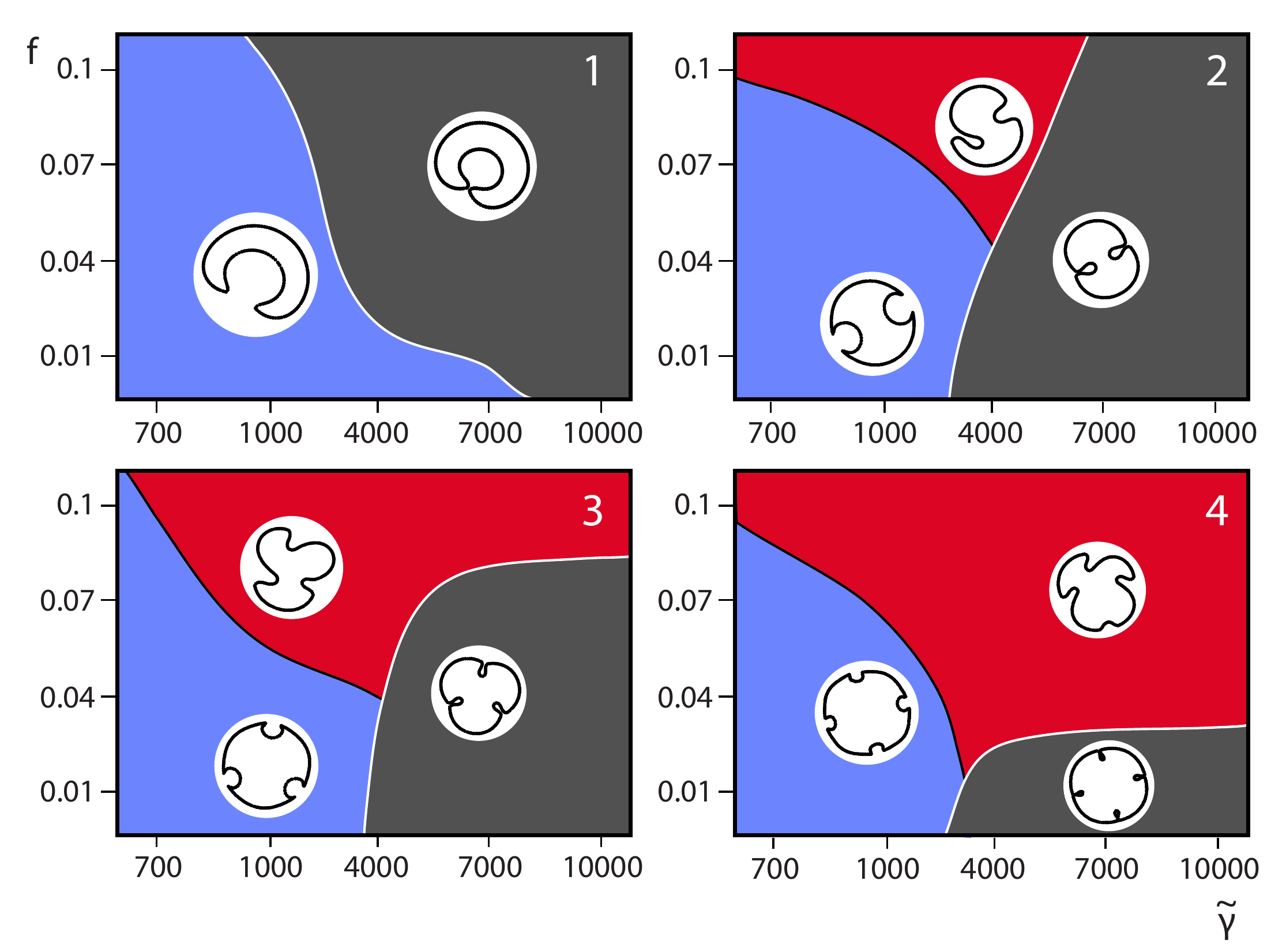}
\caption{Phase diagrams of the three main types of infolded pollen shapes in the $(f,\tilde{\gamma})$ parameter space for grains with $n=1$, $2$, $3$, and $4$ colpi (as denoted in the diagrams), with $A_{ap}/A_0=1/3$ in all cases. Light blue regions correspond to shapes in which the apertures infold symmetrically but do not close. In the red regions, the exine is significantly deformed and the apertures either close partially or only some of them close fully, and the infolding pattern is either asymmetrical or of lower symmetry than the initial shape. In the dark grey regions, all apertures completely close in a symmetrical or nearly symmetrical fashion. Each region additionally contains a representative equatorial cross-section, positioned at its approximate $(f,\tilde{\gamma})$ coordinates.}
\label{fig:4}
\end{figure}

The area covered by the apertures can also be arranged in different ways: it can be united in a single wide colpus, or distributed among several ($n$) colpi, each one covering an area of $A_c=A_{ap}/n$. The number of colpi (and apertures in general) can vary significantly among and even within pollen species. It is thus of importance to investigate how this influences the folding pathways of pollen grains. The results are summarized in Fig.~\ref{fig:4}, which shows phase diagrams of infolded pollen shapes in the $(f,\tilde{\gamma})$ plane for grains with $n=1$, $2$, $3$, and $4$ apertures. Analysis presented earlier in Fig.~\ref{fig:3} suggests that infolded shapes and pathways can be divided in roughly three categories: grains with apertures that infold symmetrically but do not close (light blue regions in the diagrams in Fig.~\ref{fig:4}), grains which infold with pronounced deformation of the exine and in which apertures do not close symmetrically and simultaneously (red regions), and grains in which all apertures completely close in a symmetrical or nearly symmetrical fashion (dark grey regions). This categorization of infolded shapes serves well not only for $n=3$, but also for $n=1$, $2$, and $4$ (cross-sections of the infolded shapes for $n=2$ and $n=4$ are shown in SI Appendix, Figs.~S5 and S6). Figure~\ref{fig:4} makes it apparent that regular folding pathways need to be more precisely elastically regulated in grains with larger numbers of apertures, seen as the decrease in the size of the dark grey regions in the phase diagrams as $n$ increases. Only sufficiently soft apertures (small $f$) guarantee a full and symmetrical infolding of the pollen grain, and this condition becomes more stringent with large $n$. Interestingly, in the case of $n=1$, the smallest values of $f$ require grains with large enough FvK numbers ($\tilde{\gamma}\gtrsim 7000$) in order to produce apertures which close completely.

The case of inaperturate pollen ($n=0$) corresponds in our model to the limiting value of $f=1$ where the apertural and interapertural regions have the same elastic properties and the shell is thus homogeneous, irrespective of $n$. As homogeneous spherical shells undergo a symmetry-breaking buckling instability upon a change in volume~\cite{Pogorelov,Vliegenthart2011}, this means that the red region in the phase diagrams in Fig.~\ref{fig:4} must necessarily appear for all $\tilde{\gamma}$ and $n$ as $f$ approaches $1$.

\subsection*{Differences in folding pathways}

\begin{figure*}[t]
\centering
\includegraphics[width=1.3\columnwidth]{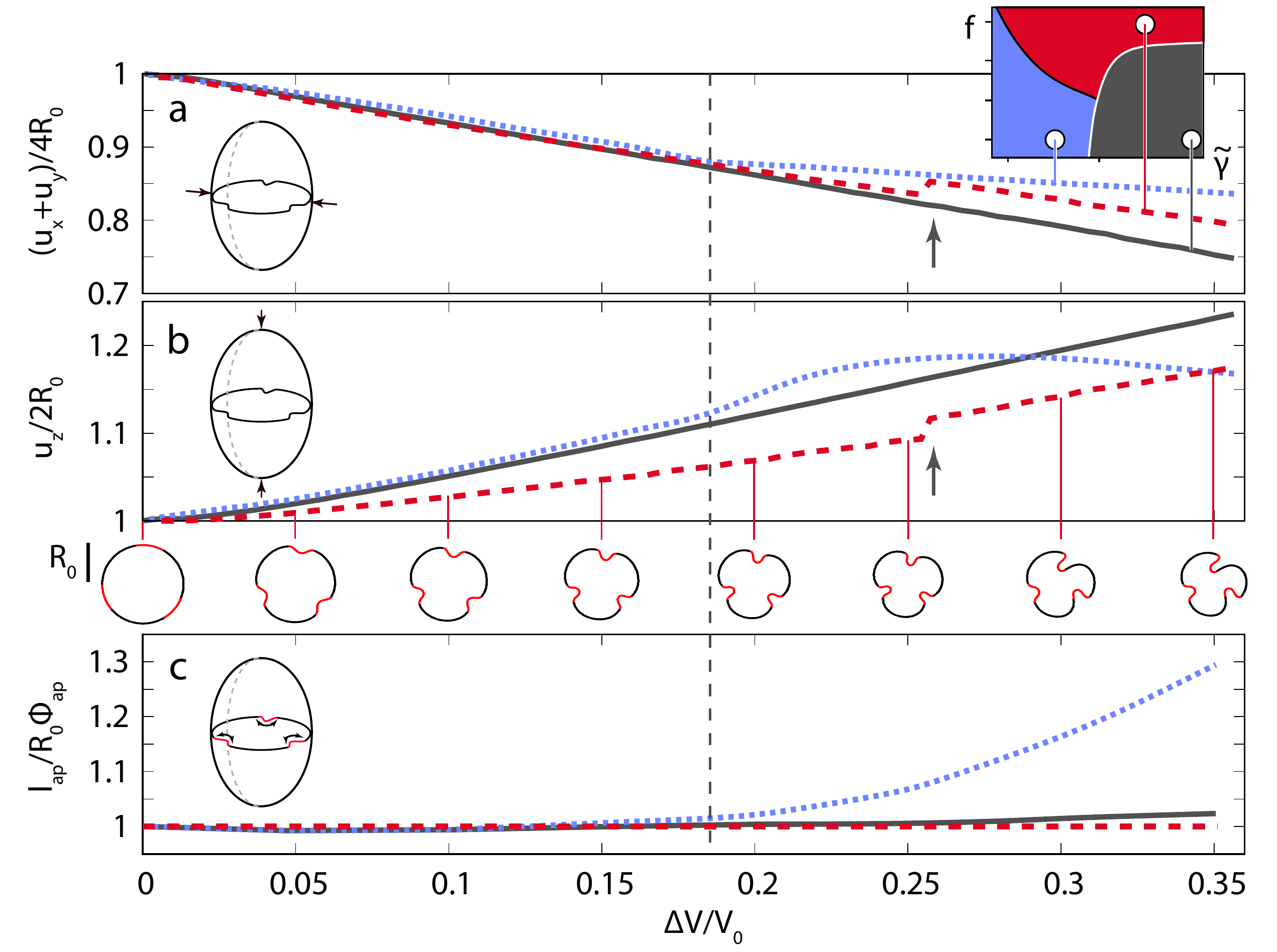}
\caption{Different folding pathways for pollen grains with $\tilde{\gamma} = 10000$, $f=0.01$ (full lines), $\tilde{\gamma} = 1000$, $f=0.01$ (dotted lines), and $\tilde{\gamma} = 7000$, $f=0.1$ (dashed lines) are illustrated by examining the geometry of the grains as they infold and $\Delta V/V_0$ increases. Shown are the change in the mean equatorial dimensions of the bounding boxes of the grain {\bf (a)}, the half-lengths of the bounding boxes in the $z$-direction {\bf (b)}, and the total lengths of the apertures in the equatorial plane {\bf (c)}. The latter were scaled with the equatorial lengths of the apertures in the fully hydrated state, $R_0 \phi_{ap}$. Insets below panel {\bf (b)} show the change in the equatorial cross-sections of a grain with $\tilde{\gamma} = 7000$ and $f=0.1$ as its volume decreases. Vertical dashed line shows the reduced volume where the folding pathways of the two grains with $f=0.01$ begin to show marked differences. Colored inset in panel {\bf (a)} shows the positions of the cases represented by the three lines in the $(f,\tilde{\gamma})$ plane of the $n=3$ phase diagram from Fig.~\ref{fig:4}.}
\label{fig:5}
\end{figure*}

To examine in more detail the differences in the calculated pollen folding pathways, we show in Fig.~\ref{fig:5} the changes in pollen shape during desiccation for three cases of tricolpate pollen grains in the three different regions of the phase diagram in Fig.~\ref{fig:4}: $\tilde{\gamma} = 10000$ and $f=0.01$ (symmetrical infolding where the apertures close; dark grey region in Fig.~\ref{fig:4}), $\tilde{\gamma} = 7000$ and $f=0.1$ (apertures do not close symmetrically and simultaneously; red region), and $\tilde{\gamma} = 1000$ and $f=0.01$ (symmetrical infolding but the apertures do not close; light blue region). Changes in the pollen shape brought about by the decrease in volume are represented by the variations of the bounding box dimensions of the grains, with Figs.~\ref{fig:5}a and \ref{fig:5}b showing the changes in the equatorial plane and along the $z$-direction, respectively. Figure~\ref{fig:5}c further shows the changes in the total lengths of the apertures in the equatorial plane, scaled by their corresponding lengths in the fully hydrated pollen. Each of the folding pathways presented in Fig.~\ref{fig:5} is a sequence of pollen grain shapes with the lowest elastic energy at each point in a series of progressively decreasing volumes.

The folding pathways of the two cases with soft apertures ($f=0.01$; dotted light blue and full dark grey lines in Fig.~\ref{fig:5}) are initially similar as the volume starts to decrease and apertures gradually change their curvatures. At about $\Delta V / V_0 \approx 0.18$, the pathways start to markedly differ and the grain with $\tilde{\gamma} = 1000$ (dotted light blue line) almost stops shrinking in the equatorial plane as its volume is reduced further. It also stops elongating along the $z$-axis, after a short acceleration of elongation when $0.2 < \Delta V / V_0 < 0.23$. This corresponds to a development of two ``pinches'' at the poles of the grain, as seen in the lower left 3D shape in Fig.~\ref{fig:3}. In the case of the grain with $\tilde{\gamma} = 10000$ (full dark grey line), no transitions in the change of the bounding box dimensions are observed as it continuously shrinks in the equatorial plane and elongates along the $z$-axis, and the apertures infold and fully close at around $\Delta V / V_0 = 0.35$. Even though the bounding box dimensions of the grain with $\tilde{\gamma} = 1000$ almost stagnate when $\Delta V / V_0 > 0.18$, the grain continues to reduce its volume. Figure~\ref{fig:5}c shows that the apertures of this shape start to inflate towards its interior. This process reduces the volume but requires the apertures to stretch, elongating in the equatorial plane by $30\%$ when $\Delta V / V_0$ increases from $\approx0.18$ to $\approx0.35$. No stretching of the apertures is observed in the case of the grain with $\tilde{\gamma} = 10000$, and this shape reduces its volume by exine bending and a complete closing of the apertures which, once fully closed, take up a much smaller interior volume when compared to the case with $\tilde{\gamma} = 1000$ ($\Delta V / V_0 = 0.35$). Large strains are typical for highly desiccated grains with small $\tilde{\gamma}$ (e.g., with thick walls), which suggests that a nonlinear elastic model may be required in order to more realistically model such grains in a regime of large deformations (see SI Appendix for details).

The irregular folding pathway pertaining to stiffer apertures or weaker exine (red dashed lines in Fig.~\ref{fig:5}) is very different from the other two, as can also be observed by inspecting the cross-sections of the infolded shapes from this pathway and comparing them with the shapes in Fig.~\ref{fig:2} (regular closing). The geometry of the pollen shape now changes non-monotonously during infolding. Both shape descriptors in Figs.~\ref{fig:5}a and \ref{fig:5}b feature a kink at $\Delta V / V_0 \approx 0.253$, indicated by an arrow. At this transition point, the C$_3$ rotational symmetry around the $z$ axis is broken, and past this transition, each of the three apertures continues to infold in a different fashion. In general, the irregular folding pathways of the red regions in the phase diagrams in Fig.~\ref{fig:4} are more prone to be influenced by noise, and the absolute minima of energy are more difficult to determine compared with the shapes where the apertures close symmetrically (see SI Appendix for a detailed discussion). The minimization procedure must thus be carried out carefully in order to determine {\em the} solution pathway rather than {\em some} deformation pathway~\cite{Vliegenthart2011}, which additionally signifies a failure of the apertures to strictly guide the infolding pathways of pollen grains whose elastic properties fall into the red regions of the phase diagrams.

The different folding pathways in Fig.~\ref{fig:5} can be explained by considering the energetics of the infolding. A complete infolding of the grain apertures requires some bending of the exine regions as well, which adopt a smaller effective radius in the equatorial plane upon infolding (see Fig.~\ref{fig:2}). Bending deformation of the exine is also required for the elongation of the grain. The exine regions must thus deviate from their spontaneous curvature, characteristic of the fully hydrated state of the grain. While the contribution of stretching in the exine deformation is small, exine bending becomes prohibitively expensive energy-wise for sufficiently small $\tilde{\gamma}$ (i.e., for grains where the ratio $Y/\kappa$ is small). Consequently, instead of a complete infolding of the apertures and the accompanying bending of the exine, the pollen grain reduces the volume by stretching (inflating) its apertures towards its interior. This decreases the bending energy of the exine and lowers the total elastic energy of the grain for small $\tilde{\gamma}$ and sufficiently large $\Delta V / V_0$. The process, however, requires very soft apertures (small values of $f$) and does not take place when the apertures are stiff enough---in that case, we typically observe an irregular infolding (red regions in Fig.~\ref{fig:4}), as the apertures are not soft enough to localize the deformation of the pollen wall.

\section*{Discussion}

Having in mind the various mechanical influences to which pollen grains are exposed on their journey from the anther to the stigma, a robust mechanical design of the grain infolding patterns that can resist various forces and preserve the regularity of the folding pathway should provide an evolutionary advantage. According to our simulations, both the overall elastic properties of the pollen grain (as given by its FvK number $\tilde{\gamma}$) as well as the softer regions provided by the apertures (and described by the softness parameter $f$) play a key role in determining the folding pathway of pollen grains upon desiccation. For instance, soft apertures are required to close the grain symmetrically---but too soft apertures will not pull in and bend the exine, and will instead yield to stress and stretch in order to reduce the volume. This will not close the grain and could result in a rupture in materials which cannot sustain large stretching deformations ($\gtrsim30\%$). On the other hand, if the apertures are too hard, they will fail to act as sites of localized deformation but will instead induce a deformation of the entire grain. If the apertures are to guide the pollen folding reliably, a delicate elastic balance is thus required which depends not only on the softness of the apertures $f$, but also on the elasticity of the entire grain, i.e., its FvK number. Small differences in the elastic properties of pollen grains are thus one possible explanation for the observed differences in folding pathways of pollen grains which otherwise appear practically the same~\cite{Halbritter2004,PALDAT}.

Folding pathways of pollen grains can be furthermore drastically influenced by the shape, size, and number of apertures. It is perhaps not surprising that circular pores do a much worse job at guiding pollen folding along a regular pathway compared to elongated colpi, as has already been observed before~\cite{Katifori2010,Payne1972}, while the area covered by the apertures appears to mainly influence the achievable reduction in the volume of the grain before the apertures close. The largest influence, however, on the available folding pathways of pollen grains seems to be provided by the number of apertures: the region of elastic parameters where desiccation leads to a regular, complete closing of the apertures shrinks with the increasing number of apertures (Fig.~\ref{fig:4}). While there is a benefit to many apertures arising from simple geometrical considerations---the probability of a grain falling on a stigma with one of its apertures touching it increases for grains with more apertures, approaching $100\%$ for sufficiently large $n$~\cite{Furness2004}, and experiments have observed a positive correlation between germination speed, number of apertures, and pollen water content at dispersal~\cite{Dajoz1991,Franchi2011}---the majority of angiosperm pollen has either one or three apertures, indicating that there must be factors selecting against too large a number of apertures as well. Our work shows that a possible reason favoring smaller numbers of apertures could be the requirement for a robust mechanical design of pollen grains, where the apertures close completely and in a regular fashion and thus slow the rate of water loss. 

The results of our study thus contextualize and develop the conclusions of the work by Katifori et al.~\cite{Katifori2010}. While they have demonstrated the role of both compliant apertures as well as exine ornamentation in achieving predictable and reversible pollen infolding, our work shows that such a pathway is realized only for a sufficiently tuned mechanical design of the grain, involving a complex interplay of pollen elasticity and the aperture number, shape, and size. A change in any of these parameters can push the pollen grain out of the dark grey regions of the phase diagram in Fig.~\ref{fig:4}---which were of primary interest in Ref.~\cite{Katifori2010}---and induce an irregular infolding of the grain (see also SI Appendix for additional information). This has been implicitly demonstrated in Ref.~\cite{Dobritsa2009}, where exine-deficient mutants of tricolpate pollen grains of {\em Arabidopsis thaliana}, exhibiting a significantly weakened exine, no longer closed properly upon desiccation despite the presence of the apertures. A weakened exine corresponds to larger values of $f$, which again shows that such a change will push the pollen grains away from a regular closing pathway. To connect these observations with our results in a more straightforward fashion, studies aimed specifically at connecting the elastic properties of different pollen to their folding pathways are needed.

The results presented in this work emphasize the importance of a more detailed theoretical and experimental exploration of the mechanical and elastic properties of pollen grains in order to obtain a better understanding of their folding pathways. The different types of pathways we have observed should be valid for any size class of pollen, as the general range of FvK numbers we considered depends on the ratio of the pollen size and the exine thickness, and larger pollen grains tend to have thicker exines~\cite{Lee1978}, thus constraining the range of FvK numbers. Given the ample variation of pollen types, however, it would not be surprising to find pollen grains which fall outside of this range, and our thin shell model should be applied with caution to the cases where the thickness of the exine becomes comparable to the size of the pollen grain. Our model should also be applied with some care to pollen grains with pronounced sculpting of the exine or the presence of ornamentation in the apertural region which can, for instance, lead to steric effects influencing the closure of the apertures~\cite{Katifori2010}. We have also demonstrated that a more complete modelling of the grain infolding will likely need to account for nonlinearity of the elastic response of the grain and allow for a more nuanced difference between the elasticity of the exine and the apertures, which is in our model accounted for by a single softness parameter $f$.

In addition to providing a mechanical explanation for some evolved features of pollen grain structure, our work should prove useful in the design of inhomogeneous elastic shells which respond reversibly to changes in the osmotic pressure. Colloidal capsules, for example, can sustain an external osmotic pressure up to a critical point after which they buckle. This process can be strongly influenced by structural inhomogeneities in the capsule shells~\cite{Datta2012}, which is consistent with our results, since apertures can be also thought of as designed inhomogeneities---these can thus dramatically change the buckling pathway of the capsule and guide it in a desired direction. Pollen grains themselves can be used as templates for containers for microencapsulation of compounds and as drug delivery vehicles~\cite{Atwe2014,Mundargi2016,Uddin2020}, and it is of interest to devise strategies to manipulate pollen beyond its natural performance limits~\cite{Fan2020}. One way to achieve this, suggested by the results of our study, is through a manipulation of the elastic properties of the exine and the apertures. This could be done for instance by gene manipulation~\cite{Dobritsa2009,Ariizumi2011,Shi2015,Quilichini2015,Wang2018} or environmental stimuli~\cite{Mercado1997,Hinojosa2019}, which can influence the thickness of the pollen wall, its sporopollenin content, or even the exine pattern, all of which has the potential to change the elastic response of the grain.

\section*{Materials and Methods}

\subsection*{Elastic model of pollen folding and desiccation}

We assume that the fully hydrated, spherical shape of the pollen grain has no residual elastic stresses in the wall. The energy of the system $E$ is constructed by formulating the discrete bending and stretching energy contributions on a spherical mesh of triangles approximating the fully hydrated grain (more details regarding the mesh construction are discussed in SI Appendix). The stretching contribution is implemented along the edges $i$ of the mesh, penalizing edge extensions and shortenings $l_i$ from their value in the fully hydrated state $l_i^0$. The bending contribution is implemented between pairs of triangles $I$, $J$ sharing an edge, so that the local bending energy is parametrized in terms of the deviations of the angles between the triangle normals $\theta_{I,J}$ from their value in the fully hydrated state $\theta_{I,J}^0$:
\begin{equation}
E = \sum_i \frac{\epsilon_i}{2} \left(l_i - l_i^0 \right)^2 + \sum_{I>J} \rho_{I,J} \left[ 1- \cos(\theta_{I,J} - \theta_{I,J}^0) \right].
\label{eq:el_energy}
\end{equation}
The stretching and bending elastic parameters ($\epsilon_i$ and $\rho_{I,J}$, respectively) mirror the elastic inhomogeneities in the grain structure, and their values are set depending on whether the edges $i$ and faces $I,J$ are located in the aperture regions ($\epsilon_{ap}, \rho_{ap}$) or in the surrounding exine regions ($\epsilon_{ex}$, $\rho_{ex}$). (Cases when edge endpoints or neighboring faces are in different domains are explained in SI Appendix.) There are obviously many different ways to set up the inhomogeneity of elastic properties: in order to make the problem tractable, we introduce a scaling, softness parameter $f<1$~\cite{Katifori2010} such that $\epsilon_{ap}=f \epsilon_{ex}$ and $\rho_{ap} = f \rho_{ex}$. A smaller value of $f$ simulates a larger elastic difference between the aperture and the exine, while $f=1$ represents the limiting point where the aperture becomes the same as its surrounding in terms of its elasticity.

Pollen desiccation is modelled by gradually decreasing the volume of the grain, treating the volume enclosed by the elastic shell $V$ as a mechanical constraint~\cite{Siber2006}. The relative change of the grain volume is given by $\Delta V/V_0$, where $V_0$ is the volume of the fully hydrated pollen and $\Delta V = V_0 - V$. Dehydration in pollen varies to a great extent~\cite{Franchi2002,Payne1972,Payne1981,Nepi2001}, and the water content of pollen correlates with the volume of the grain. In {\em Cucurbita pepo}, for example, pollen volume has been observed to be $\sim 20\%$ higher than the hydration status of the grain~\cite{Nepi2001} (meaning that the volume of the grain with, e.g., 50\% water content is about 70\% of the volume it has at 100\% water content: $V=0.7 V_0$ and $\Delta V/V_0 = 0.3$ in our notation).

\subsection*{Continuum thin-shell elasticity}

In the continuum theory of thin shells, their elastic behavior is governed by a single dimensionless parameter, the FvK number $\gamma=(Y/\kappa) R_0^2$~\cite{Lidmar2003}, which depends on the ratio of two elastic parameters: 2D Young's modulus $Y$ and bending rigidity $\kappa$; $R_0$ is the radius of the shell. Continuum elastic moduli $Y$ and $\kappa$ are proportional to the microscopic stretching and bending parameters $\epsilon$ and $\rho$~\cite{Lidmar2003}, and the constants of proportionality are on the order of one. The precise relationship between the discrete and continuum elastic constants depends, however, both on the nature of the triangulation and the geometry of the shape being triangulated, as demonstrated in Refs.~\cite{Guckenberger2017,Tsubota2014}. For our purposes, we define the quantity
\begin{equation}
\tilde{\gamma} \equiv \frac{\epsilon}{\rho} R_0^2,
\end{equation}
which we call the FvK number, having in mind that it may differ from the continuum value of the FvK number $\gamma$ by a multiplicative constant on the order of one. The exact value of this constant is irrelevant for our purposes, since the important effects take place in the interval spanning orders of magnitude in $\gamma$ and the elastic constants (and thus the FvK numbers) of pollen grains are largely unknown.

\subsection*{Elastic properties of pollen grains}

While recent experiments have estimated Young's modulus of different pollen grains (obtaining atypically high values in the process, $Y=10$--$16$ GPa for pollen with $15$--$40$ $\mu$m in diameter~\cite{Qu2018}), elastic constants of pollen remain otherwise poorly known. To estimate the range of $\gamma$ relevant for pollen grains, we thus combine the expressions connecting 2D Young's modulus and bending rigidity with 3D Young's modulus and Poisson's ratio to write the FvK number as~\cite{Landau,Vliegenthart2011}:
\begin{equation}
\gamma=12\,(1-\nu^2)\left(\frac{R_0}{d}\right)^2,
\end{equation}
where $d$ is shell thickness and $\nu$ is 3D Poisson's ratio. The latter can be estimated to be $\nu \approx 0.35$, a value appropriate for rigid polymers and based on the heavily cross-linked structure of sporopollenin~\cite{Qu2018}. While $\nu$ can vary from $0.2$ (very rigid) to $0.5$ (rubber-like), the exact number is not particularly important for our estimate of $\gamma$, since the ratio $R_0/d$ plays a more dominant role.

Even though pollen comes in sizes from less than 10 $\mu$m to more than 100 $\mu$m in diameter---depending also on the degree of hydration and the preparation method---the majority of monad, spheroidal pollen is in the range of 10--100 $\mu$m in diameter~\cite{PALDAT}. In general, pollen thickness increases with pollen size; several studies have estimated the thickness of the exine wall in different types and species of pollen to be in the range $d\in[0.5,2.5]$ $\mu$m~\cite{Matamoro2016,Bolick1990,Lee1978}. Our analysis of the available data in the online palynological database PalDat~\cite{PALDAT} indicates that for a large sample of spheroidal pollen with radii $R_0\in[5,50]$ $\mu$m an exine thickness of $d=1$--$2$ $\mu$m is a very good approximation. This excludes any pronounced variations in the exine ornamentation and sculpturing in the form of, e.g., spikes or other protrusions on the scale of $R_0$. We can thus estimate that $R_0/d\approx5$--$30$ (similarly to what has been reported in the literature~\cite{Bolick1990,Lee1978}), and that consequently the FvK numbers of pollen can span a large range, $\gamma\approx10^2$--$10^4$. Figure~\ref{fig:1} shows an example of pollen grain from {\em Betonica officinalis}, with an estimated $R_0/d\sim15$--$20$ and $\gamma\sim2$--$4\times10^3$.

Exine ornamentation on a scale much smaller than $R_0$ is not important for the process of harmomegathy, which takes place on the spatial scale comparable to $R_0$. Such effects can be thus included in some average sense in the elastic parameters of the exine. While the exine is very strong (either due to its thickness or its physical properties)~\cite{Bolick1992,Qu2018}, recent AFM studies estimate that aperture regions are at least an order of magnitude softer, and the smallest measured values of aperture Young's modulus are two orders of magnitude smaller than the measured Young's modulus of exine~\cite{Edlund2016,Wang2018}. We can thus estimate that the values of the softness parameter $f$ are in the range of $f=0.01$--$0.1$, which is what we use in our analysis.

\subsection*{Data availability}

All the necessary data and the methods required to reproduce the results of this study are given in the main text and SI Appendix. Extensive data on grain shapes in different elastic regimes is available from the authors upon reasonable request.

\begin{acknowledgments}
AB acknowledges funding from the Slovenian Research Agency ARRS (Research Core Funding No. P1-0055).
\end{acknowledgments}

\bibliography{references}

\end{document}


\title{Mechanical design of apertures and the infolding of pollen grain:\\ SI Appendix}

\author{An\v ze Bo\v zi\v c}
\affiliation{Department of Theoretical Physics, Jo\v zef Stefan Institute, Jamova  39, 1000 Ljubljana, Slovenia}
\author{Antonio \v Siber}
\affiliation{Institute of Physics, Bijeni\v{c}ka cesta 46, 10000 Zagreb, Croatia}
\email{antonio.siber@ifs.hr}

\date{\today}

\maketitle

\section*{Supplementary Methods}

\subsection*{Details of the discrete elastic model}

The spherical mesh representing a fully hydrated pollen grain is constructed using the marching method for triangulation of surfaces, described in Ref.~\cite{hartmann1998}. The method is modified so that the starting polygon is a pentagon, instead of a hexagon as in the original implementation---this yields more uniform triangulations with no visible patches consisting of predominantly hexagonally-coordinated vertices. Results presented in the manuscript were obtained using a mesh consisting of $V=8597$ vertices, $E=25785$ edges, and $F=17190$ faces; note that $F+V-E=2$, as guaranteed by the Euler formula for polyhedra. The mesh contains $1146$, $6314$, and $1134$ vertices with $5$, $6$, and $7$ nearest neighbors (joined by edges), respectively. The maximum and minimum edge lengths of the triangulation are $1.73\,a$ and $0.58\,a$, respectively, where $a$ is the average edge length. The radius of the spherical mesh is $R_0=24.34\,a$. Calculations were also performed with smaller and larger meshes, and the chosen mesh was found to be sufficiently large so that the results obtained do not depend in any important aspect on the nature and details of the triangulation.

Mesh vertices are assigned either to the apertures or to the exine region, depending on whether their coordinates fall within the mathematically defined borders of the apertures or not (see below). The same assignment is also made for the faces (triangles) of the triangulation based on the coordinates of their centroids. The elastic energy model is formulated for the edges of the mesh. A pair of microscopic stretching and bending constants ($\epsilon_i$, $\rho_i$) is assigned to each edge $i$, depending on the position of its endpoints and on the faces $I$ and $J$ it joins. If both endpoints are either in the aperture or in the exine region, the edge is assigned the stretching constant pertaining to the aperture, $\epsilon_{ap}$, or the exine, $\epsilon_{ex}$, respectively. If one of the endpoints belongs to the aperture and the other to the exine region, the edge is assigned a stretching rigidity of $(\epsilon_{ap} + \epsilon_{ex})/2$. If both faces joined by the edge belong either to the aperture or to the exine region, the edge is assigned the bending rigidity of $\rho_{ap}$ or $\rho_{ex}$, respectively. Similarly, if one of the triangle faces belongs to the aperture and the other to the exine region, the edge is assigned the bending rigidity of $(\rho_{ap} + \rho_{ex})/2$.

\subsection*{Minimization procedure}

Vertex coordinates of the shapes minimizing the energy functional are calculated by a nonlinear conjugate gradient minimization procedure described in Ref.~\cite{hager2006}. The volume constraint is implemented as an energy penalty of the form $\lambda (V_{actual} - V_{target})^2$, where $V_{actual}$ and $V_{target}$ are the actual and target volumes of the system. The constraint parameter $\lambda$ is typically sequentially increased until the volume found is within the predefined tolerance (always less than $10^{-5}$ in our calculations)~\cite{Siber2006}. Once the minimum of the energy is found, it is checked that the total contribution of the energy penalty is vanishingly small when compared with the elastic energy of the system.

Folding pathways are calculated by sequentially reducing the target volume and using the shape obtained in the previous step of the procedure as the initial guess for the conjugate gradient minimization. The procedure is started with the fully hydrated, spherical shape and its corresponding volume $V_0$. Once the minimal target volume is reached, the procedure is repeated backwards---from the maximally infolded shape towards the fully hydrated shape---by sequentially increasing the target volume. This procedure enables detection of lower energy states which may have been missed in the forward minimization but can still be detected in the backward minimization, and vice versa.

A precise mapping of the folding pathways is not essential for the main results of our work---what matters is whether the shape retains its (approximate) C$_3$ symmetry the entire way to closing or undergoes a transition at some volume prior to a complete closing of the apertures. This information can be obtained with less numerical effort than the mapping of the complete infolding pathway, i.e., the branch with the lowest energy clearly separated from other solution branches~\cite{Xia2020}. Nevertheless, it is important to investigate the stability and convergence of numerical results and the ensuing robustness of the folding phase diagrams presented in Fig.~4 in the main text. The minimum energy shapes of the grains which infold in an irregular fashion (red regions of the phase diagrams) are particularly sensitive to numerical details, and the minimization procedure must be devised with special care in order to approach {\em the} minimum energy solution rather than {\em some} solution with a higher energy. This is especially true near the borders of the regions in the phase diagrams in Fig.~4, where the details of the folding pathways (such as shown in Fig.~5 in the main text) are found to be influenced by noise.

To approach the lowest energy states, multiple minimizations with different values of the volume penalty parameter $\lambda$ were performed. A slightly less or more strict volume constraint can sample different solution branches in particularly sensitive cases with essentially the same accuracy of the target volume. To further enhance the detection of different solution pathways in order to find the one with minimum energy, we have also introduced in some calculations a jittering of the positions of all vertices between two sequential minimization steps along a folding pathway. Each of the mesh vertices was displaced randomly in a cube centered at its equilibrium position. The cube edge length was typically $1/10$ of the average edge length of the mesh $a$ or less. Finally, since the irregular folding pathways break the discrete rotational symmetry of the grain around the $z$ axis, the energy of the symmetrical infolding pathway is also tracked in these regions in order to check whether some other solution has perhaps a lower energy. This pathway is fixed by seeding the minimization with the symmetrically infolded shape from the regular region (dark grey areas in Fig.~4), typically one with a small softness parameter $f$ and the same value of $\tilde{\gamma}$ (see also the SI Methods subsection on Hysteresis of folding pathways for more details).

\subsection*{Convergence of results on meshes of different sizes}

We have verified the convergence of our numerical results by performing shape minimizations on meshes of different sizes, i.e., with different numbers of triangular faces. The number of faces in a mesh $N_f$ is proportional to $(R_0/a)^2$---a very accurate estimate is given by $N_f = 16 \pi (R_0/a)^2  / \sqrt{3}$. We constructed four meshes with $(R_0/a)^2$ = 422.3, 532.3, 592.4, and 1044.6 ($N_f$ = 12254, 15446, 17190, and 30312); the mesh with $(R_0/a)^2 = 592.4$ is the one used in all calculations described in the main text. The volume enclosed by the mesh is an explicit parameter of the calculation, and the maximum volume considered, measured in units of $a^3$, is almost four times larger than the minimum mesh volume considered.

\begin{figure}
\centering
\includegraphics[width=0.55\textwidth]{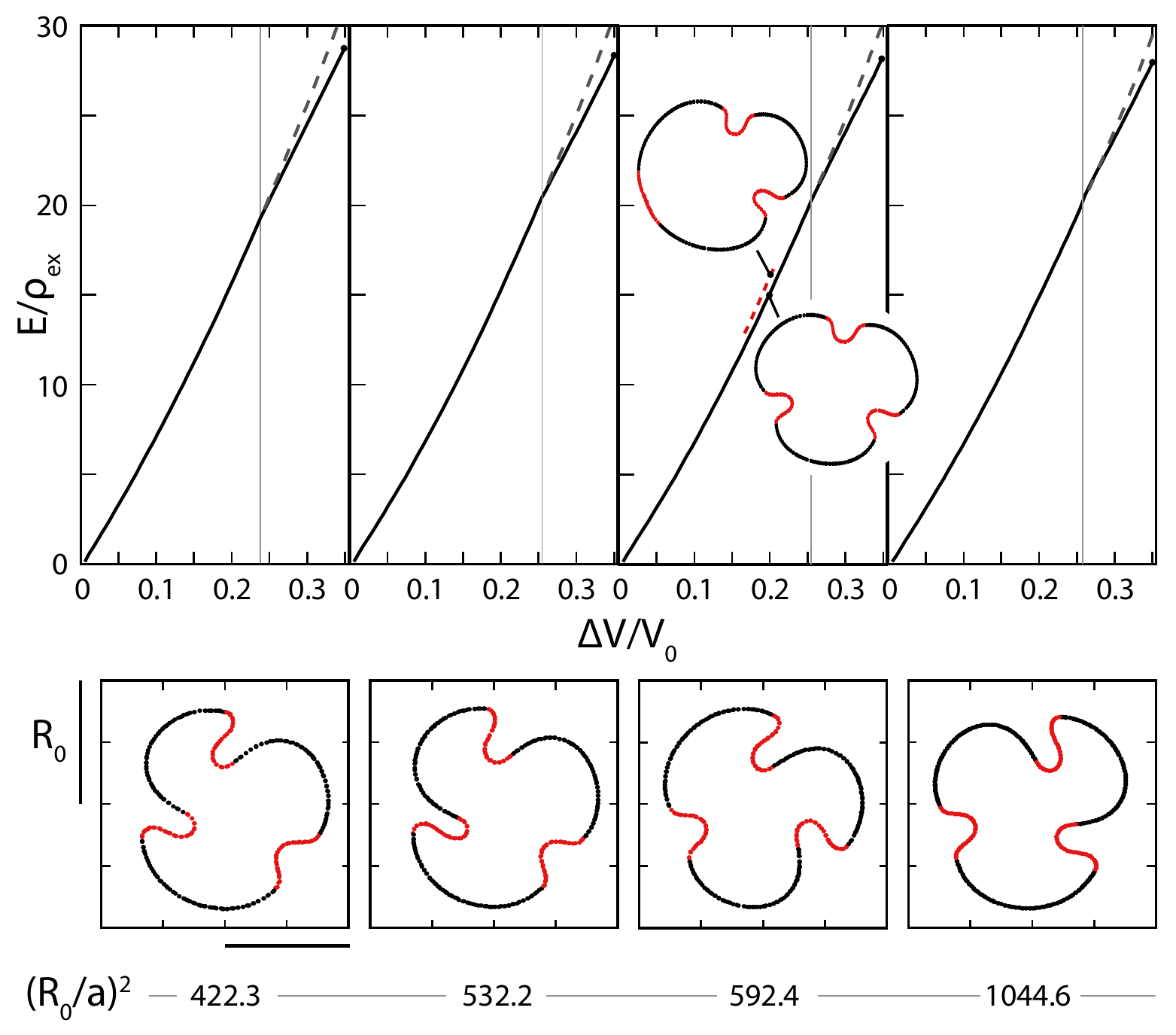}
\caption{Top row: elastic energies of infolding pathways (full lines) of a pollen grain with $\tilde{\gamma}=7000$ and $f=0.1$. The calculations were performed on spherical meshes of four different resolutions, as indicated by the squared ratio of the mesh radius and the average length of the mesh edge $(R_0/a)^2$. Symmetry-breaking transition points are denoted by thin vertical lines. Dashed grey lines show elastic energies of the shapes which retain C$_3$ symmetry past the transition points. Short-dashed red line shows a branch with a larger energy for $(R_0/a)^2=592.4$. For the same mesh, insets show two equatorial cross-sections corresponding to shapes with $\Delta V/V_0 = 0.2$ and belonging to either the higher energy branch or the pathway branch, as indicated by arrows. Bottom row: equatorial cross-sections of the minimum energy shapes with $\Delta V/V_0 = 0.35$ obtained on different mesh resolutions as indicated. Red and black regions in the equatorial cross-sections denote the apertures and the exine, respectively.}
\label{fig:convergence1}
\end{figure}

In Fig.~\ref{fig:convergence1}, we show the folding pathways for a pollen grain with $\tilde{\gamma}=7000$ and $f=0.1$ (dashed red line in Fig.~5), calculated for each of the four meshes. Top row of panels in Fig.~\ref{fig:convergence1} shows the minimum energies determined at each reduced volume (full lines) as well as the corresponding energies of shapes with C$_3$ symmetry (dashed grey lines). We can observe a symmetry-breaking transition at a specific reduced volume of the grain (indicated by a thin vertical line), after which the solutions with the C$_3$ symmetry no longer represent the minimum energy branch. To obtain these higher energy branches, the folding pathway was started at a small value of $\Delta V / V_0$ (typically $\Delta V / V_0 \approx 0.1$) and the shape with C$_3$ symmetry that was used as the initial guess was taken from the minimization of a grain with softer apertures, $\tilde{\gamma}=7000$ and $f=0.01$, at the same reduced volume. The pathway was then evolved towards larger values of $\Delta V / V_0$ without any noise---such a procedure is often found to preserve the C$_3$ symmetry of the shapes determined by the conjugate gradient minimization, even past the transition point. The solution with C$_3$ symmetry is the minimum energy solution below the transition point $\Delta V / V_0 \approx 0.25$, the exact value of which is only very slightly dependent on the choice of mesh size.

The bottom row of panels in Fig.~\ref{fig:convergence1} shows the minimum energy shapes for the four different mesh sizes, calculated at $\Delta V / V_0 = 0.35$, a value near aperture closing for shapes with small $f$ (Fig.~3 in the main text). The energies of these shapes are $E/\rho_{ex} = 28.8$, $28.3$, $28.2$, and $28.0$ as mesh resolution increases, and we see that they become progressively smaller---which is to be expected of a calculation with desirable convergence properties. Despite some differences, the obtained shapes are very similar to each other: this is obviously so for the two leftmost shapes obtained at mesh sizes $(R_0/a)^2 = 422.3$ and $532.3$, and the shape at $(R_0/a)^2 = 592.4$ needs to be rotated counterclockwise by $120^\circ$ in order to visually align with the first two shapes. When reflected across the line passing through the center of the shape and the middle of the symmetrically infolded aperture, the rightmost shape at $(R_0/a)^2=1044.6$ also reduces to the shape at $(R_0/a)^2 = 592.4$. All the shapes are thus characterized by very similar curvatures and energy, and any differences in the shapes are almost nullified under appropriate symmetry transformations, a consequence of the degeneracy of the solutions reflecting the C$_3$ symmetry of the fully hydrated grain. This is a restricted version of the effect which appears in the minimization of a homogeneous sphere under pressure, where a depression can occur in an infinite number of equivalent positions~\cite{Pogorelov,Vliegenthart2011}.

\subsection*{Hysteresis of folding pathways}

Folding pathways sometimes exhibit hysteresis: the shape obtained in backward minimization (the pathway of increasing grain volume) is not the same as the one obtained in the forward minimization (the pathway of decreasing grain volume). Typically, the shape obtained in backward minimization has a lower elastic energy, but this is not always the case; this implies multiple stable solution pathways. When such situations occur in our minimization, this often indicates that the pathway leads to irregular closing and that there is a transition point somewhere along the pathway, which can be determined more precisely using the techniques described previously (seeding with the appropriate symmetry, introducing noise, repeating the calculations with different volume penalization parameters). The shapes obtained on a higher energy branch of the hysteresis curve do not belong to the minimum pathway but might have some relevance for the grain infolding. A part of the higher energy solution branch is shown by a dashed red line in Fig.~\ref{fig:convergence1}. The shape obtained along this branch, shown in the inset of the energy plot for $(R_0/a)^2$ = 592.4, does not have a C$_3$ symmetry characteristic of the minimum shape shown in the inset below it (for $\Delta V / V_0 = 0.2$); it does, however, retain a mirror symmetry. Interestingly, such a shape is not found in the infolding pathway neither prior to the transition at $\Delta V / V_0 \approx 0.25$, nor past it---it always remains a solution with higher energy, but its detection can serve to indicate the irregular nature of the infolding.

\subsection*{Model of the aperture shape} 
\begin{figure}
\centering
\includegraphics[width=0.55\textwidth]{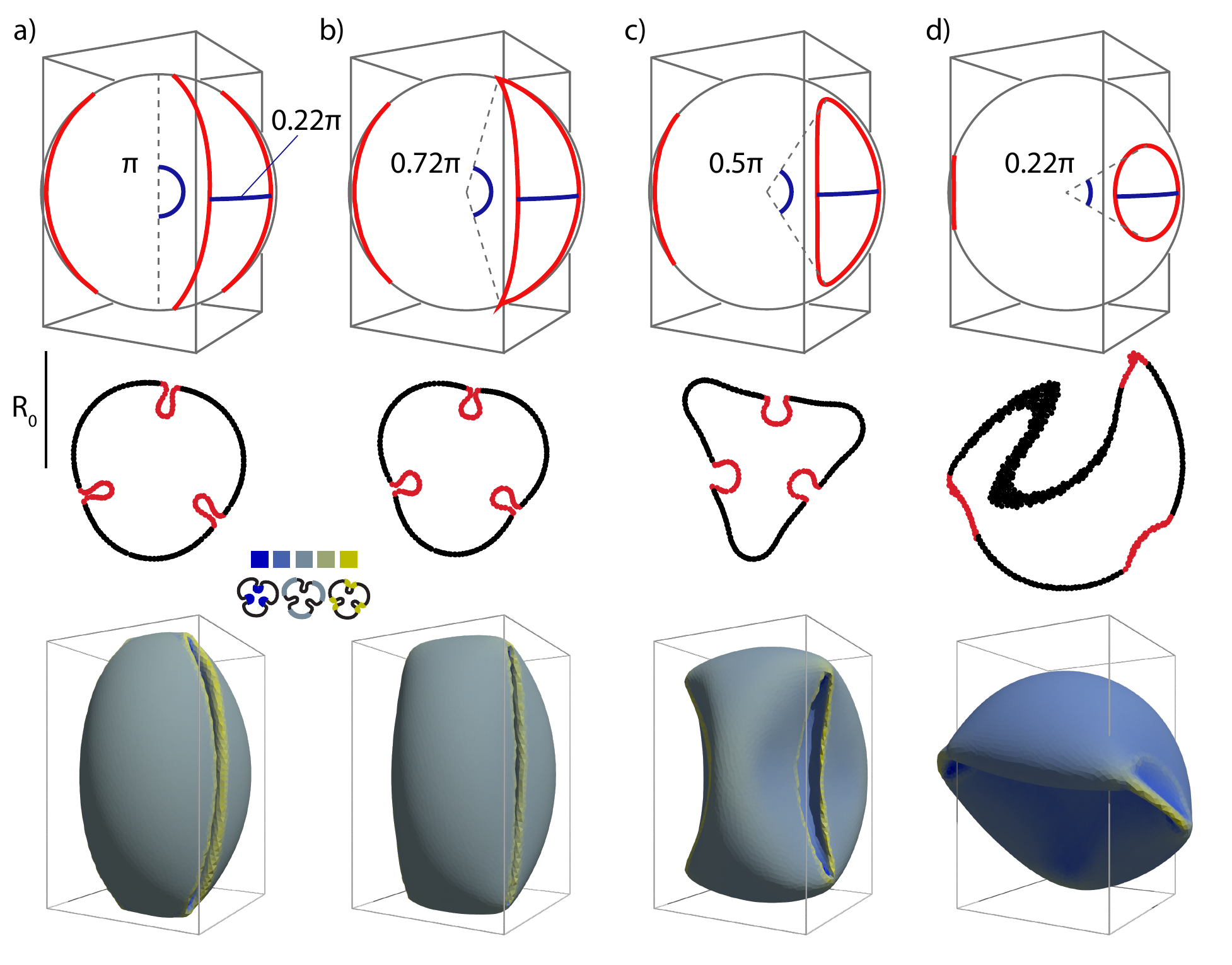}
\caption{Shapes of desiccated triaperturate grains with $\tilde{\gamma}=7000$ and $f=0.01$ at $\Delta V/V_0=0.35$ for different aperture sizes and shapes. The span of the aperture along the azimuthal angle is the same in all four cases, $\phi_c=0.22\pi$, while the span along the polar angle is gradually reduced from $\theta=\pi$ (pole-to-pole colpus) to $\theta=0.22\pi$ (circular pore). This means that the total proportion of the aperture area $A_{ap}/A_0$ is gradually reduced as well---see the main text and SI Methods for more details. Colors in 3D shapes indicate the mean surface curvature, with the brightest yellow and the darkest blue corresponding to largest positive and negative curvatures, respectively.}
\label{fig:S1}
\end{figure}

Apertures are arranged equatorially, and the angular width of each aperture along the equator is $\phi_c$. The total equatorial angular span of all $n$ apertures is $\phi _{ap} = n \phi _c$. Centres of neighboring apertures are separated by an azimuthal angle of $\alpha = 2 \pi / n$, where $n$ is the number of apertures in the pollen grain. The azimuthal angle $\varphi \in [0,2\pi]$ is rescaled so that $\tilde{\varphi} = \varphi - {\rm int}(\varphi / \alpha)\,\alpha$, where ${\rm int}(x)$ denotes the integer part of $x$. All points for which the azimuthal angle $\tilde{\varphi}$ and the polar angle $\vartheta \in [0,2\pi]$ fulfill the conditions
\begin{eqnarray}
&\tilde{\varphi}& < \quad \phi_c \nonumber \qquad{\rm and} \\
\frac{\pi}{2} - \theta_{\Psi}(\tilde{\varphi}, \phi_c) \quad <  &\vartheta& < \quad \frac{\pi}{2} + \theta_{\Psi}(\tilde{\varphi}, \phi_c)
\end{eqnarray}
belong to the aperture. The apertures are arranged along the equator of the grain, where they assume the largest azimuthal width. We have also defined $\theta_{\Psi}(\tilde{\varphi}, \phi_c) = \Psi \sqrt{\left(\frac{\phi_c}{2}\right)^2  - \left(\tilde{\varphi} - \frac{\phi_c}{2} \right)^2}$, where the parameter $\Psi$ determines the shape of the aperture. When $\Psi = 1$, the aperture is approximately circular (porus). As $\Psi \rightarrow \infty$, the aperture assumes a shape of a spherical lune (idealized colpus). The maximal polar span of the aperture in this case, $\theta=\pi$ (pole-to-pole), is obtained when $\tilde{\varphi} = \phi_c/2$. For finite values of $\Psi$, the maximal polar span is given by $\theta={\rm max}[\Psi \phi_c, \pi]$. Calculations presented in the main text were performed with $\Psi \rightarrow \infty$. Figure~\ref{fig:S1} shows equatorial cross-sections (middle row) and 3D shapes (bottom row) of infolded triaperturate grains for different shapes of the apertures (from an idealized colpus to porus; top row).

\section*{Minimum volume and the prolateness of maximally infolded shapes}

\subsection*{Bounding box dimensions}

Dimensions of the bounding box of a pollen grain are calculated with respect to fixed coordinate axes. This means that the bounding box should not be considered as the minimal bounding box in some coordinate-independent sense (e.g., minimum volume or minimal equatorial area box). While different definitions of the bounding box can produce somewhat different ratios of the box dimensions, any difference would be small in our case, as the cross-section of the grain in the equatorial plane is nearly isometric, i.e., it typically does not show a prominent elongation along a particular direction in the equatorial plane.

\subsection*{Analytical approximations for the bounding box dimensions and the volume of the maximally infolded shape}

\begin{figure}
\centering
\includegraphics[width=0.55\textwidth]{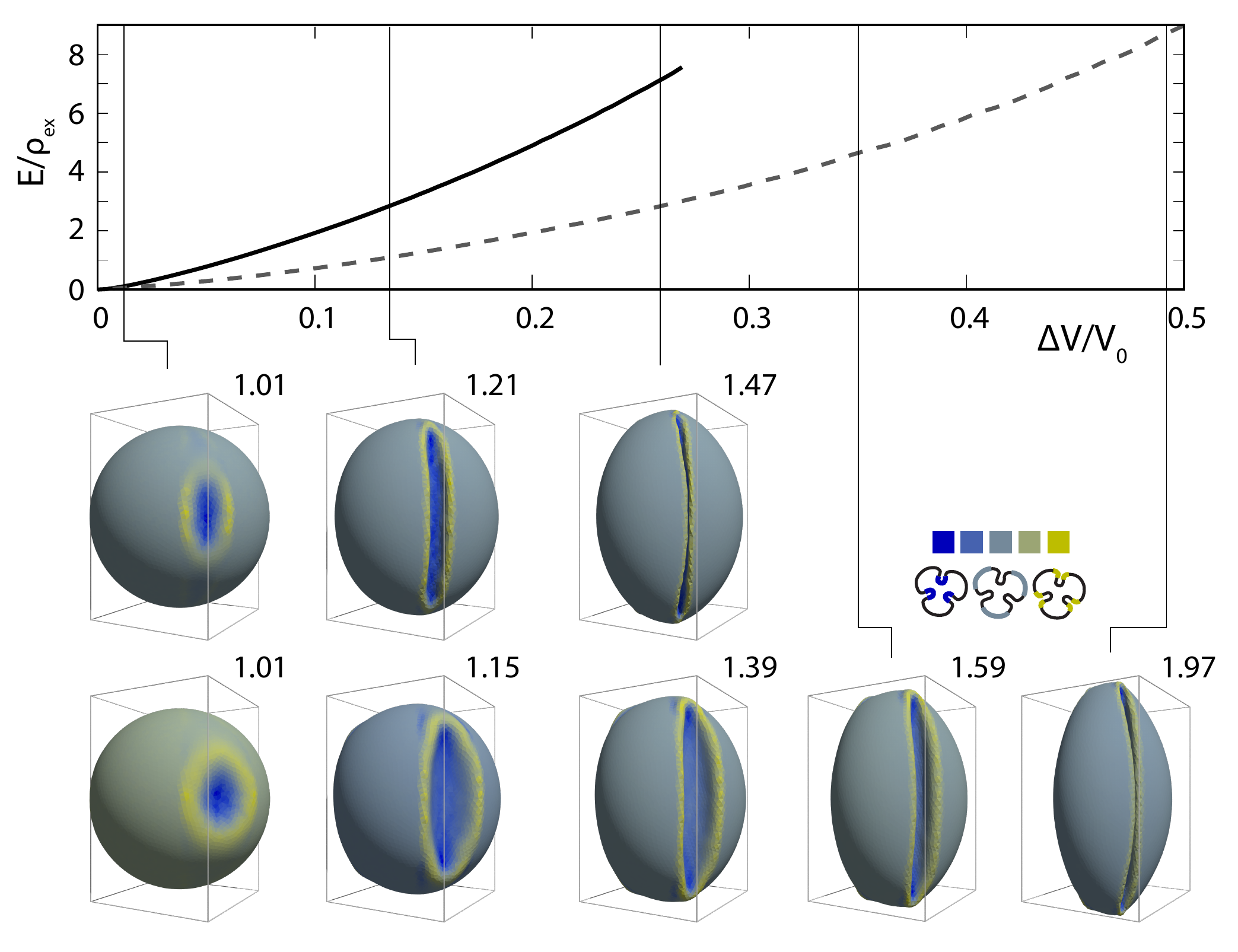}
\caption{Regular aperture closing during desiccation of tricolpate pollen with $\tilde{\gamma}=7000$, $f=0.01$, and $A_{ap}/A_0=0.25$ (black line) and $0.41$ (dashed grey line). The elastic energy of the shapes is shown as a function of volume reduction $\Delta V/V_0$ (top panel). The characteristic shapes of the shells are shown in the middle and bottom rows of graphs for $A_{ap}/A_0=0.25$ and $0.41$, respectively, for $\Delta V/V_0=0.012$, $0.13$, $0.26$ (middle row) and $\Delta V/V_0=0.012$, $0.13$, $0.26$, $0.35$, $0.49$ (bottom row). Colors in 3D shapes indicate the mean surface curvature, with the brightest yellow and the darkest blue corresponding to largest positive and negative curvatures, respectively.}
\label{fig:S2}
\end{figure}

The final, minimum volume of a fully infolded shape, $V_{in}$, is a quantity which depends on the total area of the apertures $A_{ap}$, their number $n$, and, in general, also on the elasticity of the grain. A rough estimate of $V_{in}$ can be obtained by assuming that the exine of the infolded shape does not stretch significantly and by approximating the shape of the grain by an ellipsoid. The cross-section in the equatorial plane can be approximated by a circle whose perimeter can be calculated by formally excising the apertures. The longest axis of the ellipsoid can be obtained from the requirement that the exine does not stretch when the equatorial radius of the grain shrinks. From these considerations, one can calculate the dimensions of the bounding box of the grain at the point of complete infolding as
\begin{eqnarray}
\frac{u_x+u_y}{4 R_0} &=& 1 - \tilde{\phi}_{ap} \nonumber \\
\frac{u_z}{2 R_0} &=& 1 + \tilde{\phi}_{ap},
\end{eqnarray}
where $\tilde{\phi}_{ap} = \phi _{ap} / 2\pi = A_{ap} / A_0$. For $\tilde{\phi}_{ap} = 1/3$, this gives $(u_x+u_y)/(4 R_0) = 2/3$, and $u_z/(2 R_0) = 4/3$, which compares favorably with the values of $0.75$ and $1.25$ obtained in the numerical simulations (Fig.~2 in the main text). The volume of the grain at the point of maximal infolding is given by
\begin{equation}
\frac{\Delta V_{in}}{V_0} \approx \tilde{\phi}_{ap} + \tilde{\phi}_{ap}^2 - \tilde{\phi}_{ap}^3,
\label{eq:S_vin}
\end{equation}
where $\Delta V_{in} = V_0 - V_{in}$. This rough estimate does not account for the volume subtracted from the grain by the infolded apertures themselves. For $\tilde{\phi}_{ap} = 0.25$, $0.33$, and $0.41$, the equation predicts $\Delta V_{in}/V_0 \approx 0.30$, $0.41$, and $0.51$, respectively, in reasonable agreement with the values obtained numerically for $n=3$ ($0.27$, $0.35$, and $0.5$; see Fig.~\ref{fig:S2} and Fig.~2 in the main text). Equation~(\ref{eq:S_vin}) less accurately describes grains with $n=1$ and $2$, because the infolded apertures take up a significant volume of the infolded grain. This effect is rather difficult to estimate analytically, and we have found the following approximation to be useful:
\begin{equation}
\frac{\Delta V_{in}}{V_0} \approx \tilde{\phi}_{ap} + \tilde{\phi}_{ap}^2 - \tilde{\phi}_{ap}^3 - \frac{\tilde{\phi}_{ap}^2}{2n} ( 1 + \tilde{\phi}_{ap}) 
\end{equation}
The last term, subtracted from the approximation in~\eqref{eq:S_vin}, accounts for the volume taken from the grain by the infolded apertures. The correction is small when $n>2$, but becomes important for $n=1$ and $2$. The smallest infolded volume of a grain is obtained when the entire aperture area is lumped in a single, large aperture ($n=1$).

\section*{Elastic strains in the infolding shell}

Our calculations use a Hookean-type model, which is most easily recognized in the stretching contribution to the energy (Eq.~(1) in the main text). It is thus of interest to examine the magnitude of strains in the infolding grains predicted by the model in order to estimate the importance which nonlinear effects might play. The change of the equatorial length of the apertures, shown in Fig.~5, is a global measure of the strains. While the global strains in the apertures appear to be small along the entire folding pathway for the pollen grain with $\tilde{\gamma}=10000$ and $f=0.01$ (dark grey region in the phase diagram in Fig.~4) and for the grain with $\tilde{\gamma}=7000$ and $f=0.1$ (red region), they are quite significant for the grain with $\tilde{\gamma}=1000$ and $f=0.01$ (light blue region) for $\Delta V / V_0 > 0.2$, reaching $30\%$ at $\Delta V / V_0 = 0.35$. A complete analysis of the strains in the latter case shows that they locally reach even larger values. The maximum relative extensions and compressions of the edges $(l-l_0) / l_0$ are $0.99$ and $-0.99$, respectively, but such values are highly localized. Nevertheless, this shows that nonlinear effects might play an important role in this case, and more generally for desiccated grains with sufficiently small values of $\tilde{\gamma}$ (thick shells) and $f$.

Analysis of the strain distribution in a regularly infolding tricolpate grain with $\tilde{\gamma}=10000$ and $f=0.01$ at $\Delta V / V_0 = 0.16$ is shown in Fig.~\ref{fig:S_inelastic}. The edges of the mesh are colored according to their relative extension/compression $(l-l_0) / l_0$, as indicated by the color scale. In this case, the maximal compressions and extensions of the edges are $-10\%$ and $4\%$, respectively, and they reach $-19\%$ and $8\%$ when the volume is decreased further to $\Delta V / V_0 = 0.35$ (full infolding of the grain). Large strains are localized in the apertural region, and the largest strains (edge compressions) are even more localized in two small portions of the aperture separating the apertural depression from the poles of the pollen grain. Such a strain distribution is different from the situation in buckled homogeneous shells, where the strain is equally distributed along a ridge of a circular depression resembling the shape of an inverted spherical cap~\cite{Pogorelov}. A specific shape of the colpi induces a breaking of the axial symmetry of the depression, leading to strains which concentrate near the grain poles---quite a similar solution has also been found in Fig.~4 in Ref.~\cite{Katifori2010}. We do, however, obtain an axisymmetric strain distribution when $f=1$, i.e., when our model reduces to the description of a buckling of a homogeneous elastic shell. A cup-like shape, typical for the buckling of homogeneous spheres, is also obtained when the apertures are small and completely fail to accommodate the change in the grain volume---see, e.g., the shape in Fig.~\ref{fig:S1}d.

\section*{Maximally infolded shapes of bi- and tetraaperturate grains}

Figures~\ref{fig:S3} and~\ref{fig:S4} show phase diagrams, analogous to the one shown for tricolpate ($n=3$) pollen in Fig.~3, for pollen grains with two and four colpi, respectively. The shapes of infolded pollen, shown in these diagrams, were analyzed in order to produce Fig.~4 in the main text. We also analyzed the phase diagram for monocolpate ($n=1$) pollen; however, for the case of a single and sufficiently large aperture, the minimal infolded volume can no longer be thought of as a quantity independent of the elasticity of the problem. Representing the cross-sections of the infolded pollen shapes in a diagram for a single volume of the infolded grain, as was done for $n=2$, $3$, and $4$, can thus be somewhat misleading. In order to produce the data for $n=1$ required to produce Fig.~4, one must instead follow the folding pathway all the way to the point of minimum achievable volume or to the point where it becomes obvious that the apertures do not close at all. This analysis was performed for $n=1$ as well as for $n=2$, $3$, and 4, although the information in the latter three cases can be also reliably obtained from Figs.~3, \ref{fig:S3}, and \ref{fig:S4}.

\section*{Brief comparison with the work by Katifori et al.~\cite{Katifori2010}}

\begin{figure}
\centering
\includegraphics[width=0.55\textwidth]{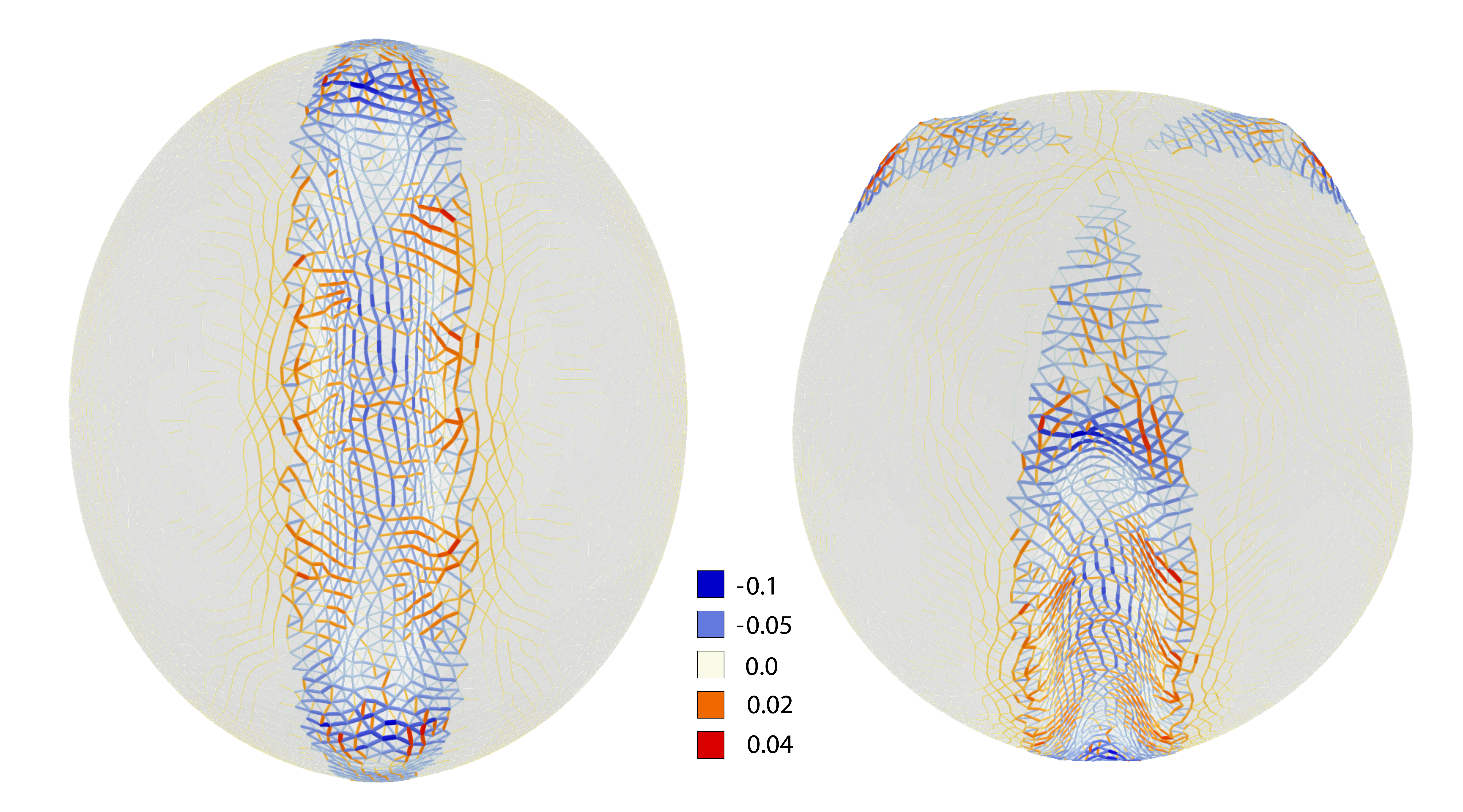}
\caption{Strains in a pollen grain with $\overline{\gamma}=10000$ and $f=0.01$ at $\Delta V / V_0 = 0.16$. The edges of the mesh are colored according to their relative extension/compression $(l-l_0)/l_0$, as indicated by the color scale. The grain is presented from two different viewpoints in order to better show the complete region of the aperture.}
\label{fig:S_inelastic}
\end{figure}

\begin{figure}
\centering
\includegraphics[width=0.55\textwidth]{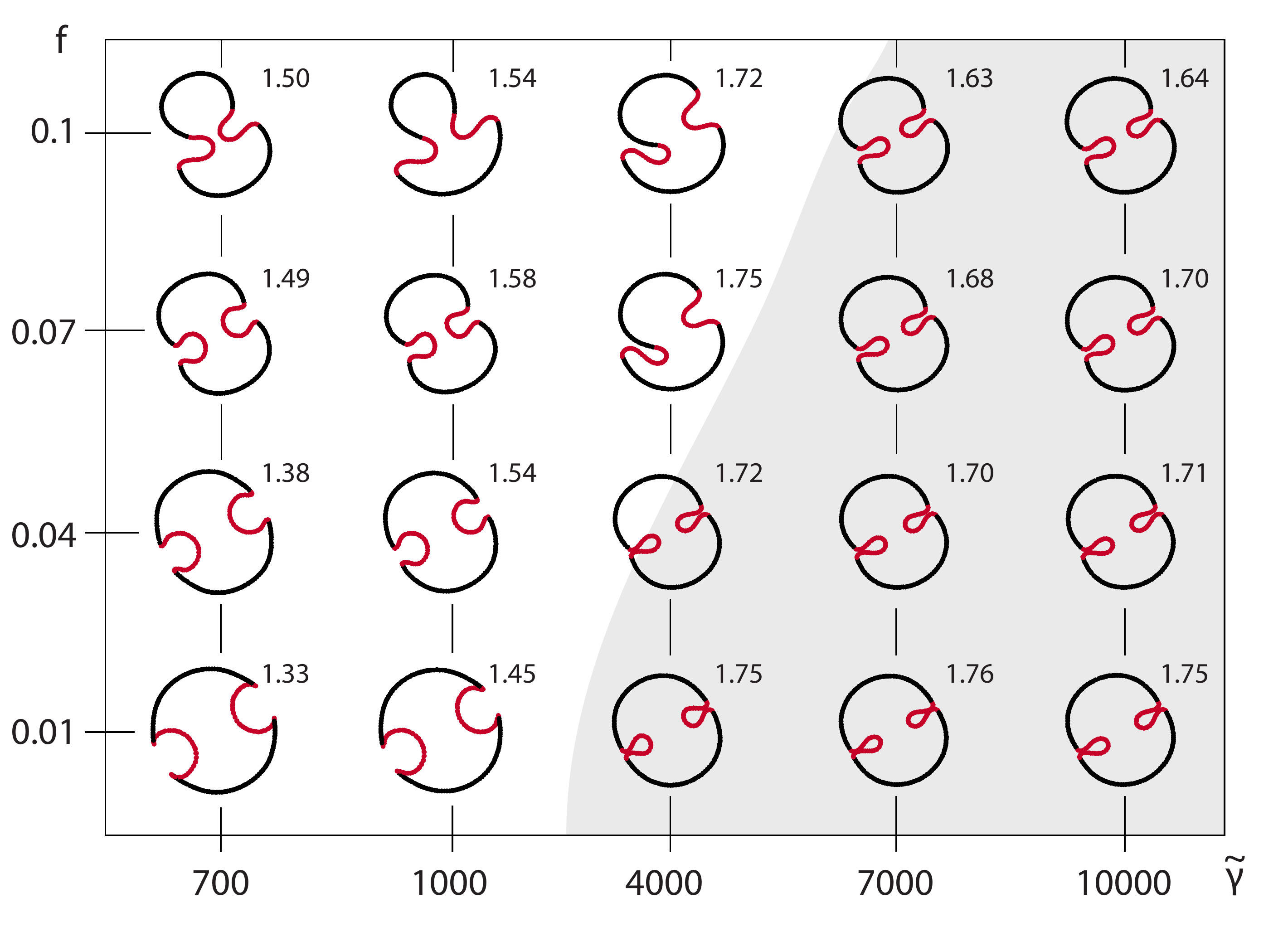}
\caption{Phase diagram of bicolpate pollen folding in the $(f,\tilde{\gamma})$ plane, showing equatorial cross-sections of the pollen shapes. The relative change of volume upon desiccation is $\Delta V/V_0=0.42$ throughout, and the colpi (shown in red in the cross-sections) span $A_{ap}/A_0=1/3$ of the total area of the pollen surface. Numbers next to the pollen shapes denote the elongation of the bounding box of the shapes, $2u_z/(u_x+u_y)$, indicating the prolateness of the grain. The shaded region of the phase diagram corresponds to the elastic parameters where the apertures close completely and nearly symmetrically, without asymmetric deformation of the exine.}
\label{fig:S3}
\end{figure}

\begin{figure}
\centering
\includegraphics[width=0.55\textwidth]{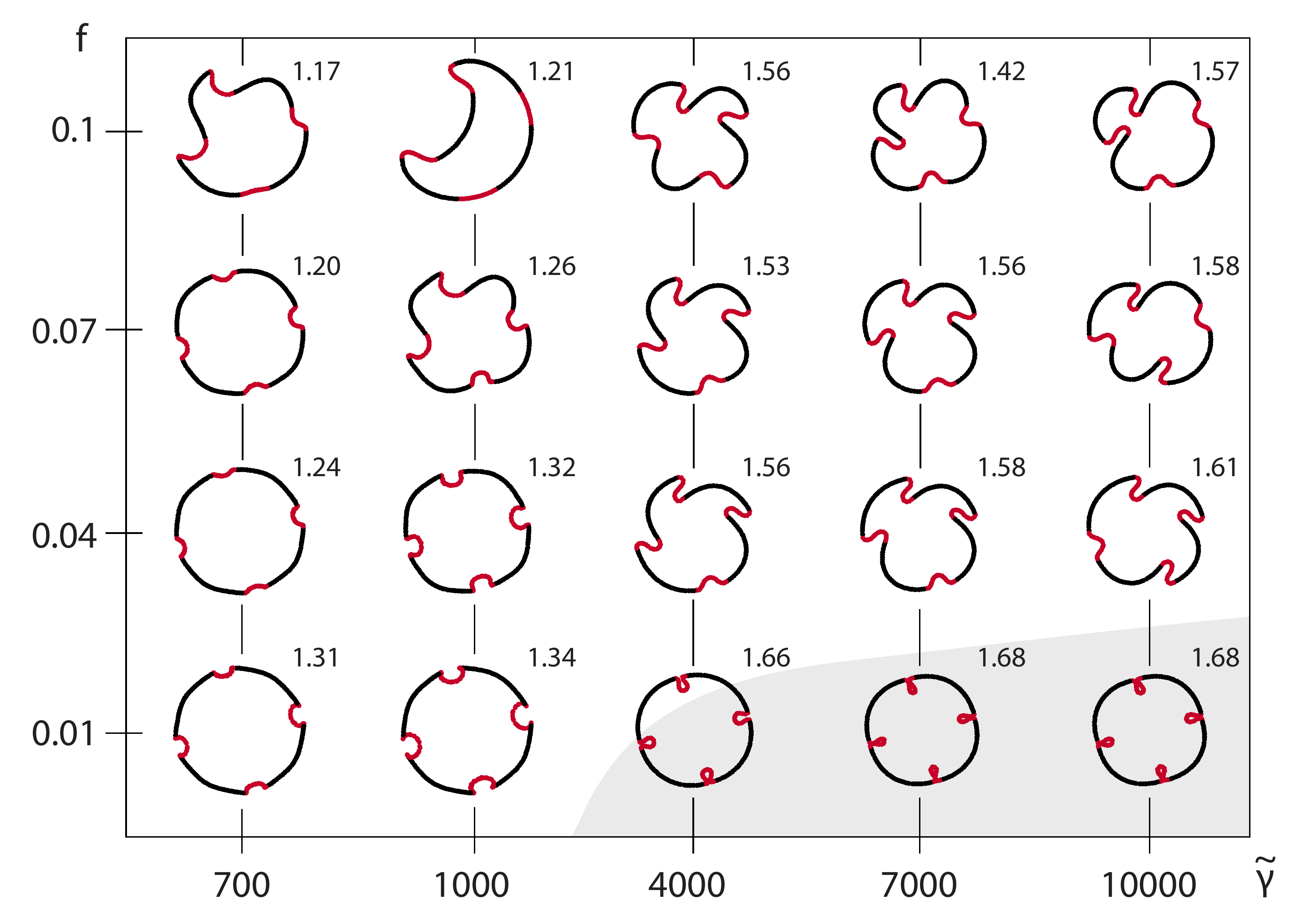}
\caption{Phase diagram of tetracolpate pollen folding in the $(f,\tilde{\gamma})$ plane, showing equatorial cross-sections of the pollen shapes. The relative change of volume upon desiccation is $\Delta V/V_0=0.35$ throughout, and the colpi (shown in red in the cross-sections) span $A_{ap}/A_0=1/3$ of the total area of the pollen surface. Numbers next to the pollen shapes denote the elongation of the bounding box of the shapes, $2u_z/(u_x+u_y)$, indicating the prolateness of the grain. The shaded region of the phase diagram corresponds to the elastic parameters where the apertures close completely and nearly symmetrically, without asymmetric deformation of the exine.}
\label{fig:S4}
\end{figure}

The elastic model used in our work is similar to the one used previously by Katifori et al.~\cite{Katifori2010} to study the role of pollen wall structure in harmomegathy, but nonetheless differs in some aspects. Their simulations were mostly performed at a single value of the ratio of the elastic parameters of the exine and the apertures---in our notation, $f=0.015$. Unlike us, however, they have also assumed that the apertures are twice as thick as the exine: an assumption which is true for some pollen, such as in the case of {\em Lilium longiflorum} they have studied, but can otherwise vary to a great extent~\cite{PALDAT} (see also Fig.~1 in the main text). The values of the ratio $R/h$ they used in their numerical model thus approximately translate to FvK numbers around $\gamma\sim10^4$, while the FvK numbers of the two polen species they focused on (monosulcate {\em Lilium longiflorum} and tricolpate {\em Euphorbia milii}) can be estimated to be $\tilde{\gamma}\gtrsim 2\times10^4$ and $\tilde{\gamma}\sim 10^3$, respectively~\cite{Katifori2010}. The fully hydrated pollen grain was modelled in their work as a prolate spheroid in order to better fit the shape of the pollen species they studied. The pollen grains which we model are perfectly spherical in the fully hydrated state, which resembles the situation in many species~\cite{PALDAT} and has the advantage of reducing to the classical problem of the buckling of a homogeneous sphere when $f=1$. The triangular mesh employed in the model of Katifori et al.~\cite{Katifori2010} is based on the triangulation of an icosahedron and has as such twelve five-fold coordinated vertices, while all other vertices are six-fold coordinated. The meshes we employ do not have any particular symmetry and appear to be more uniform. In addition to six-fold and many more five-fold coordinated vertices, they also contain a large number of seven-fold coordinated vertices (see Fig.~\ref{fig:S_inelastic}), and the numbers of the differently coordinated vertices are constricted by Euler's theorem on polyhedra. Despite all these differences, the two models yield quite comparable solutions with respect to regularly infolding grains. This can be best seen by comparing Fig.~4 in Ref.~\cite{Katifori2010} with our Fig.~\ref{fig:S_inelastic}---the spatial distributions of the stretching strains appear to be similar in the two models.

Our work demonstrates that a combination of a very low value of $f$ and a relatively high value of $\tilde{\gamma}$ leads to a complete closing of apertures in both monocolpate and tricolpate spherical pollen (Fig.~4 in the main text), which parallels the observations of Katifori et al.~\cite{Katifori2010}. The analysis we performed provides, however, a wider framework for the interpretation of these results and explains them in terms of the interplay between the elastic properties of both the exine and the apertures and the overall elasticity of the pollen grain. The results of our study additionally show that only a limited variation in the elastic properties is permitted if the grain is required to close in a regular fashion---this happens only when the FvK number $\tilde{\gamma}$ is sufficiently large and $f$ sufficiently small. The regular infolding pathway is thus not necessarily realized and can be perturbed if the elastic and geometrical properties of the grain are modified or not adjusted properly. Even when the softness of the apertures is sufficient to localize the infolding, this does not necessarily mean that the apertures will guide the infolding to a full closure of the grain. Depending on the elasticity of the exine (as given by the FvK number), the apertures can fail to pull in the exine sufficiently enough in order to seal the grain. They might instead tend to inflate into the grain interior, leaving the apertural slits open. Our results identify these three substantially different modes of infolding and relate their occurrence to the elastic properties of the pollen grain. The region of the elastic parameters leading to a regular and symmetric closing of the apertures shrinks as the aperture number increases, requiring a more precise tuning of the elastic properties of the grain. Our study thus provides a framework which can be used to contextualize the results by Katifori et al.~\cite{Katifori2010} in a wider picture of the interplay of the pollen elastic parameters, the number of apertures, and their shape and size. Such a framework also represents a basis for considerations regarding the possible evolutionary advantages of different mechanical designs of pollen grains.

\bibliography{SI-references}

\clearpage
\newpage

